\documentstyle[12pt, epsf]{article}
\def\ltsim{\lower3pt\hbox{$\, \buildrel < \over \sim \, $}}
\def\gtsim{\lower3pt\hbox{$\, \buildrel > \over \sim \, $}}
\def\be{\begin{equation}}
\def\ee{\end{equation}}
\def\ba{\begin{eqnarray}}
\def\ea{\end{eqnarray}}
\def\bea{\begin{eqnarray}}
\def\eea{\end{eqnarray}}

\def\ga{\mathrel{\raise.3ex\hbox{$>$\kern-.75em\lower1ex\hbox{$\sim$}}}}
\def\la{\mathrel{\raise.3ex\hbox{$<$\kern-.75em\lower1ex\hbox{$\sim$}}}}

\begin{document}

\baselineskip=16pt 
\begin{titlepage}
\rightline{IHES/P/02/35}
\rightline{LPT-Orsay-02/14}
\rightline{OUTP-02/06P}
\rightline{hep-th/0206042}
\rightline{June 2002}  
\begin{center}

\vspace{1cm}

\large {\bf  EFFECTIVE LAGRANGIANS
 AND UNIVERSALITY CLASSES OF NONLINEAR BIGRAVITY}

\vspace{1 cm}
\normalsize

\centerline{\bf Thibault Damour$^{\,a}$\footnote{damour@ihes.fr} and
  Ian I. Kogan\footnote{i.kogan@physics.ox.ac.uk}$^{\,a,b,c}$}
\smallskip 
\medskip 
$^a$ 
{\em IHES, 35 route de Chartres, 91440, Bures-sur-Yvette,  France }\\
$^b$
{\em Laboratoire de Physique Th\'eorique, 
%\footnote{Unite Mixte de Recherche du CNRS (UMR 8627)},
Universit\'e de Paris XI, 91405 Orsay C\'edex, France
}\\
$^c$
{\em Theoretical Physics, Department of Physics, Oxford University,
 1 Keble Road, Oxford, OX1 3NP,  UK}
\smallskip

\vskip0.6in \end{center}

\centerline{\large\bf Abstract}

We  discuss  the fully non-linear formulation of multigravity. 
The concept of universality classes of effective Lagrangians 
 describing  bigravity, which is the simplest form of multigravity, is 
introduced. We show that non-linear multigravity theories can naturally 
 arise in several different physical contexts:  brane configurations,  
 certain  Kaluza-Klein reductions and some  non-commutative geometry models.
 The formal and phenomenological  aspects of multigravity (including the 
 problems linked  to   the linearized theory of massive gravitons) are 
 briefly  discussed.
\vspace*{2mm} 
%\smallskip\newline

\end{titlepage}

\section{Introduction}

One of the most important problems which is facing theoretical physics now is 
the blending of the Standard Model (SM) with General Relativity (GR). Whatever 
way we choose (the most popular ones  nowadays are based on some 
multidimensional constructions involving extended objects), nobody doubts that it will 
definitely 
modify physics at {\it short scales}. On the other hand, the current general 
paradigm is to keep General Relativity unchanged at {\it large scales}, but to 
add new forms of gravitating matter beyond the Standard Model (dark matter, 
dark 
energy) for explaining pressing astrophysical and cosmological facts such as 
galactic rotational curves and the accelerating universe. In the present 
paper, 
we consider an alternative paradigm: a {\it modification} of General 
Relativity 
at large scales as a possible explanation of some pressing cosmological issues 
(notably cosmic acceleration).

The modification of GR that we are going to consider is linked to the issue of 
``massive gravity'' (for very light gravitons, with Compton wavelength of 
cosmological scale). A generic prediction of multidimensional constructions is 
the existence of massive gravitons. In particular, any Kaluza-Klein (KK) model 
predicts, besides a massless graviton, the presence of an infinite tower of 
massive gravitons. However, it seems impossible to use the tower of massive KK 
gravitons to modify gravity at large scales. Indeed, its spectrum is 
generically 
regularly spaced (as illustrated on Fig.~1a), so that, even if the first mode 
were very light (i.e. of cosmological Compton wavelength), there would exist 
no 
regime where the first mode (or first few modes) would be important, and where 
one could truncate away the rest of the tower of massive states. In other 
words, 
as soon as the first mode is important, we open the extra KK dimensions (see, 
however, below). The situation is, however, different in some brane models. In 
particular, Refs.~\cite{Kogan:2000wc}-\cite{Kogan:2000xc} discovered the 
possibility (illustrated in Fig.~1b or Fig.~1c) of having a {\it hierarchical 
gap}, $m_1 \ll m_2$, between the first mode (or first group, or even band, of 
modes) and the tower of higher modes. This situation, called {\it 
multigravity} (see \cite{Kogan:2001ub} for a review and \cite{Papazoglou:2001cc} for
 detailed presentation), makes it possible to envisage an effective 
 four-dimensional theory which contains only the massless 
and ultra-light gravitons and discards the states of mass $m \geq m_2$. The 
constructions \cite{Kogan:2000wc}-\cite{Kogan:2000xc} predict see-saw-like 
spectra, $m_1 \, M_{\rm Planck}^{1+\gamma} \sim m_2^{2+\gamma}$, with $\gamma$ 
interpolating \cite{Kogan:2001ub} between $0$ \cite{Kogan:2000wc} and $1$ 
\cite{Gregory:2000jc}. Such spectra are naturally compatible with the 
phenomenologically interesting situation where $m_1^{-1}$ is of cosmological 
order, while $m_2^{-1}$ is smaller than the millimetre scale.
\begin{figure}[ht]
\begin{center}
\epsfxsize=5in
\epsfbox{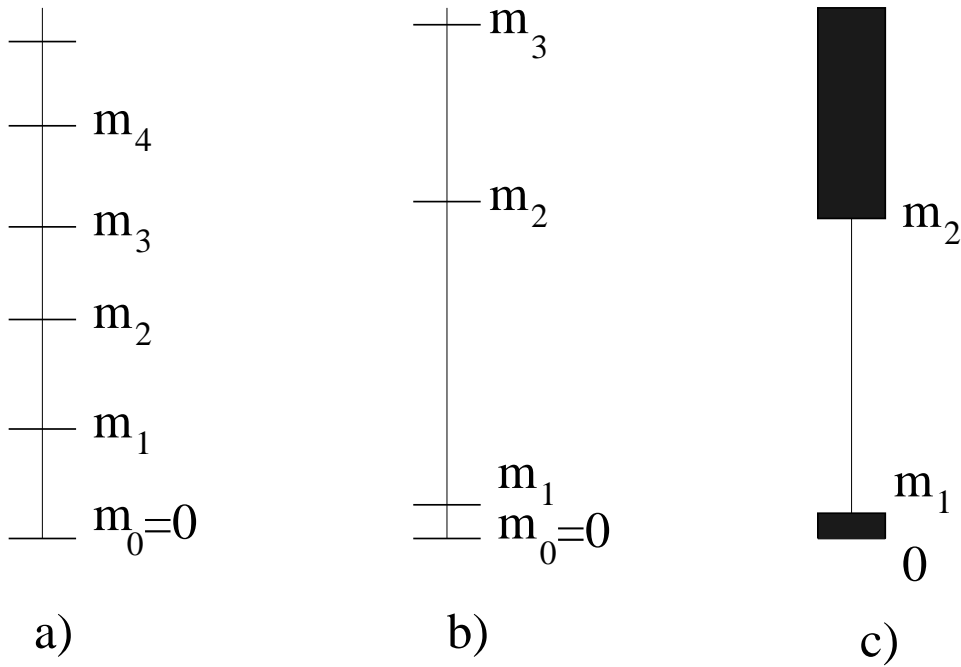}
\caption{Regular spectrum on (Fig.1 a) versus bigravity (Fig.1 b) or 
quazi-localized gravity (Fig. 1 c). The last spectrum is continuous but the 
first band is very narrow in comparison with the gap between bands.}
\end{center}
\end{figure}

So far multigravity was only analyzed in the linearized approximation. The 
main emphasis of this paper is to provide a fully non-linear formulation of 
multigravity, i.e. to write down, and analyze, a class of consistent effective 
four-dimensional Lagrangians, describing, in some limit, the light-mode 
truncation of the hierarchical spectra of Figs.~1b or 1c. Though we shall 
illustrate below our approach in the context of particular multidimensional 
realizations (notably brane models exhibiting multilocalization 
\cite{Kogan:2000wc},\cite{Kogan:2001wp}
 or  quasi-localization 
\cite{Gregory:2000jc}, \cite{Csaki:2000pp}, \cite{Dvali:2000rv}), we view our considerations as 
concerning a very general phenomenon: the concept of {\it Weakly Coupled 
Worlds} (WCW). The concept of WCW is very simple: one assumes that there are 
several Universes (labelled by $i = 1 , \ldots , N$), each endowed with its 
own metric $g_{(i)\mu\nu}$ and set of matter fields $\{ \Phi_i \}$, which are 
coupled only through some mixing of their gravitational fields. We require 
that the theory describing the WCW be near a point of enhanced symmetry, in 
the sense that there exists a limit (say as some parameter $\lambda 
\rightarrow 0$) where the theory contains $N$ diffeomorphism-like symmetries, 
corresponding to $N$ massless gravitons. A recent theorem \cite{BDGH} has 
proven that the only consistent non-linear theory involving $N$ massless 
gravitons is the sum of $N$ decoupled GR-type actions
\be
\label{eq1}
S_0 = \sum_{i=1}^{N} S [g_i , \Phi_i ] \, ,
\ee
with (we use the signature $-+++$)
\be
\label{eq2}
S [g_i , \Phi_i ] = \int d^4 x \, \sqrt{-g_i} \, [M_i^2 R (g_i) - \Lambda_i + 
L (g_i , \Phi_i)] \, .
\ee
Therefore, the only consistent action for a theory of worlds coupled only 
through gravity is of the form
\be
\label{eq3}
S_{\rm tot} = \sum_{i=1}^{N} S [g_i , \Phi_i ] + \lambda \, S_{\rm int} (g_1 , 
g_2 , \ldots , g_N) \, .
\ee
When $\lambda \rightarrow 0$, the $N$ worlds are non interacting (which 
implies that, from the point of view of any observer in one world, the other 
worlds have only a meta-physical existence), and the theory has the enormous 
symmetry $\Pi_i {\rm Diff}_{(i)}$, where each diffeomorphism group ${\rm 
Diff}_{(i)}$ acts separately on its own metric $g_{(i)\mu\nu}$ and matter 
fields $\{ \Phi_i \}$. In the interacting case, $\lambda \ne 0$, the symmetry 
of the full action must (again because of the theorem \cite{BDGH}) be reduced 
to (at most) one group of diffeomorphisms: the diagonal group of common 
diffeomorphisms transforming all metrics as
\be
\label{eq4}
\delta \, g_{\mu \nu}^{(i)} = \epsilon^{\lambda} \, \partial_{\lambda} \, 
g_{\mu \nu}^{(i)} + \partial_{\mu} \, \epsilon^{\lambda} \, g_{\lambda \nu}^{(i)} 
+ \partial_{\nu} \, \epsilon^{\lambda} \, g_{\mu \lambda}^{(i)} \equiv 
D_{\mu}^{(i)} \, \epsilon_{\nu} + D_{\nu}^{(i)} \, \epsilon_{\mu} \, .
\ee
This symmetry restricts the interaction term $\lambda \, S_{\rm int} (g_1 , 
\ldots , g_N)$ to depend only on the invariants one can make with several 
metrics. This even leaves room for extra kinetic terms built from covariant 
derivatives such as $g_{(i)}^{\mu \nu} \, D_{\lambda}^{(j)} \, g_{\mu 
\nu}^{(k)}$ (such terms do not exist in the case of one metric because 
$D_{\lambda}^{(i)} \, g_{\mu \nu}^{(i)} \equiv 0$). However, in view of the 
many potential diseases associated to modifications of the standard 
Einsteinian kinetic terms, and in the spirit of describing the class of 
interaction terms most relevant at large scales,\footnote{See Section 2.1
 below for further discussion about extra kinetic terms.} 
i.e. containing the lowest possible number of derivatives ( namely zero, 
as expected from a generalization 
of the mass terms that appear in linearized multigravity), we shall only 
consider ultra-local interaction terms, i.e.
\be
\label{eq5}
\lambda \, S_{\rm int} = -\mu^4 \int d^4 x \, {\cal V} (g_1 (x) , \ldots , 
g_N (x)) \, ,
\ee
where $\mu$ is a mass scale (henceforth replacing $\lambda$ as ``small 
parameter'') and where ${\cal V}$ is a scalar density made out of the values 
of the $N$ metrics at the same ``point''. We assume, for simplicity, that the 
$N$ weakly coupled worlds ``live'' on the same abstract manifold, i.e., in 
other terms, that one is given a family of (smooth) canonical one-to-one maps: 
${\rm world}_{(i)} \rightarrow {\rm world}_{(j)}$.

 The aim of this paper is 
threefold: (i) to motivate the possibility of the effective action 
(\ref{eq3}), (\ref{eq5}) by considering several different specific models 
(brane models, Kaluza-Klein models and non-commutative geometry ideas); (ii) 
to delineate and parametrize the various ``universality classes'' of 
non-linear multigravity; and (iii) to sketch the main qualitative consequences 
of such non-linear multigravity theories and to contrast them with the usual 
paradigm of ``massless plus massive gravitons'' which is based on a linearized 
approximation.

It should be noted that theories defined by (\ref{eq3}), (\ref{eq5}) (in the 
``bigravity'' case: $N=2$) were first introduced in the seventies \cite{ISS} 
as a model for describing a sector of hadronic physics where a massive 
spin-2 field 
(the ``$f$ meson'', with ``Planck mass'' $M_f \sim 1 \, {\rm GeV}$ 
in Eq.~(\ref{eq2})) plays a dominant role. It was then called ``strong 
gravity'' or the ``$f$-$g$ theory''. Our work not only proposes to revive, 
within a new (purely ``gravitational'') physical context, this early proposal, 
but initiates the task of systematically studying the general phenomenological 
consequences of the action (\ref{eq3}), (\ref{eq5}). The present paper will 
only briefly sketch the new physical paradigm following from such actions. In 
subsequent papers, we shall discuss in detail the cosmological consequences of 
such theories \cite{DKP1}, as well as its strong-field phenomenology 
\cite{DKP2}.

\section{Universality Classes of Bigravity Effective \\ Lagrangians}

For simplicity, we focus, in this paper, on the  case of 
``bigravity'', i.e. $N=2$. Understanding this case is a prerequisite for 
understanding the general multigravity case ($N>2$). Let us note also that the 
bigravity ``potentials'' that we discuss here can be immediately used in the 
general case.  Indeed  a rather general class of ``$N$-metric potentials'' 
${\cal V}(g_{(1)} , \ldots , g_{(N)})$, Eq.~(\ref{eq5}), is the class containing only 
``two-metric interactions'': ${\cal V} = \sum_{i \ne j} {\cal V} (g_{(i)} , 
g_{(j)})$. For instance, one can define a ``crystal-like'' many-world with 
``nearest neighbour'' interactions only ${\cal V} = \sum_i {\cal V} (g_{(i)} , 
g_{(i+1)})$. It is interesting to note that the continuum limit 
($N \rightarrow \infty$) for some suitable ``nearest neighbour'' interactions
 can mimic the propogation of gravity in a higher-dimensional space, 
i.e. the term $ \int dy a^{4} (y)
 \, \sqrt{-g(x,y)} 
 \left[ {\rm tr} (g^{-1} \, \partial_y \, g)^2 - ({\rm tr} \, 
g^{-1} \, \partial_y \, g)^2 \right] \, .
$  as in  (\ref{eq3.4}) below. See \cite{DKP1} for further discussion of 
 this  subject.
\subsection{Parametrization of invariants}

Using, when $N=2$, the notation $g_{(1)} = g_L$ (for ``Left'') and $g_{(2)} = 
g_R$ (for ``Right''), and factoring a conventional ``average volume factor'' 
$(g_L \, g_R)^{1/4}$ out of the scalar density ${\cal V}$
\footnote{We could, instead, have factored out of ${\cal V}$ the other natural symmetric
 density: $\sqrt{-g_L} +  \sqrt{-g_R}$.}, the generic 
bigravity action reads
\bea
S = \int d^4 x \, \sqrt{-g_L} \left( M_L^2 \, R(g_L) - \Lambda_L \right)  + 
\int d^4 x \, \sqrt{-g_L} \, L(\Phi_L, g_L) + \nonumber \\
\int d^4 x \, \sqrt{-g_R} \left( M_R^2 \, R(g_R) - \Lambda_R \right)  + 
\int d^4 x \, \sqrt{-g_R} \, L(\Phi_R, g_R) \nonumber \\
- \mu^4 \int d^4 x \left( g_R \, g_L \right)^{1/4} V (g_L , g_R) \, .
\label{eq6}
\eea
Note that the bigravity action (\ref{eq6}) contains 5 dimensionfull parameters: 
two ``Planck masses'' $M_L$ and $M_R$, two cosmological constants $\Lambda_L$, 
$\Lambda_R$ (with dimensions $M^4$), and the ``coupling mass scale'' $\mu$.

Before proceeding, we note that the mass scale $\mu$, entering 
Eq.~(\ref{eq6}), will be treated here as a constant parameter determining the 
coupling of the two worlds. However, one should keep in mind the possibility 
that it be replaced by a fluctuating field. This is suggested, in particular, 
by the brane realizations of bigravity where the value of $\mu$ depends on the 
physical distance between the branes, which is controlled by dilaton/radion 
fields. A more general model where $\mu \rightarrow \mu (x)$, and where one 
adds a kinetic term for $\mu (x)$, may play an important role in addressing 
crucial cosmological issues (such as inflation) in the context of multigravity 
theories.

Before we shall proceed further let us make two additional comments

\begin{itemize}
\item{We shall  treat $V$ as a potential here, i.e. as an ultra-local 
function of $g_L$ and $g_R$. As already mentioned above one could also 
 include extra kinetic terms like 
$ g^{\mu\nu}  D^{L,R}_{\mu} g_{R,L}^{\sigma\rho}
D^{L,R}_{\nu} g_{L,R~\sigma\rho}$, etc. For example  mixed terms
 like $\sqrt{-g_L}R(g_R)$ or $\sqrt{-g_R}R(g_L)$ are of these type.
 However, let us emphasise again that the fundamental concept of WCW explored 
 here is that one is required to be  near the point of enhanced symmetry 
 and according to the theorem proven in \cite{BDGH} in  this point 
 mixing of different metrics is forbidden. Because of this 
 all higher derivative terms  mixing  different metrics must be supressed
 by the small parameter $\mu^4$ and   enter  as $\mu^4 
f( D^2 g/ M^2)$. One can see  that  in the long-wave limit $D^2 g/ M^2 \sim
 k^2/M^2 << 1$ one can neglect 
their  contribution in comparison with a potential term. }

\item{Let us stress that we did not introduce any direct 
 coupling between  the  two worlds exept the indirect one mediated 
 by  gravity. It means that all matter which can be
 observed  by any observer is locally coupled only to 
 one world or  the other  one. Thus any local 
experimental check of the equivalence principle is the same as in 
the General Relativity - all locally mutually observable matter moves 
in the same metric. One can consider  other cases but such a study 
  will be left for future publications.}
\end{itemize}

The common diffeomorphism invariance (\ref{eq4}) restricts the scalar 
potential $V(g_L , g_R)$ entering Eq.~(\ref{eq6}) to depend only on the 
invariants of the mixed tensor $H \equiv g_L^{-1} \, g_R$, i.e.
\be
\label{eq7}
H_{\nu}^{\mu} \equiv g_L^{\mu \sigma} \, g_{R \sigma \nu} \, .
\ee
In 4 dimensions, there are (because of Cayley's theorem) only 4 independent 
scalar invariants which can be made from $H$. For instance, using a matrix 
notation for $H$, one can take the first 4 traces of the powers of the matrix 
$H$, say
\be
\label{eq8}
\tau_n \equiv {\rm tr} (H^n) \, ; \qquad n = 1,2,3,4 \, .
\ee

Let us introduce the 4 eigenvalues $\lambda_a$ ($a = 0 , \ldots , 3$) of 
$H_{\nu}^{\mu}$, i.e. the 4 eigenvalues of the metric $g_R$ with respect to 
$g_L$. They can be defined either by $\tau_n = \Sigma_a \, \lambda_a^n$, or by 
writing the two metrics in a special bi-orthogonal vierbein $e_{\mu}^a$ such 
that
$$
g_{\mu\nu}^L = - e_{\mu}^0 \, e_{\nu}^0 + e_{\mu}^1 \, e_{\nu}^1 + e_{\mu}^2 \, 
e_{\nu}^2 + e_{\mu}^3 \, e_{\nu}^3 \, ,
$$
\be
\label{eq9}
g_{\mu\nu}^R = - \lambda_0 \, e_{\mu}^0 \, e_{\nu}^0 + \lambda_1 \, e_{\mu}^1 \, 
e_{\nu}^1 + \lambda_2 \, e_{\mu}^2 \, e_{\nu}^2 + \lambda_3 \, e_{\mu}^3 \, 
e_{\nu}^3 \, .
\ee

It is easily seen that, apart from an exceptional case (where two eigenvalues 
coincide, and correspond to a null eigenvector), it is generically possible to 
write Eq.~(\ref{eq9}), though maybe with a complex vierbein $e_{\mu}^a$. Indeed, 
two (but at most two) eigenvalues, say $\lambda_0$, $\lambda_1$ (one of which 
necessarily corresponds to a time-like direction) can become complex. We shall 
focus on the case where the 4 eigenvalues $\lambda_0$, $\lambda_1$, $\lambda_2$, 
$\lambda_3$ are real and positive. As we shall only deal with symmetric 
functions of the eigenvalues, this restriction is mainly a notational 
convenience which can be relaxed by analytic continuation. It is then 
convenient to parametrize the invariants of $H = g_L^{-1} \, g_R$ by means of 
the logarithms of the eigenvalues of $H$:
\be
\label{eq10}
\mu_a \equiv \ln \, \lambda_a \ ; \quad \lambda_a \equiv e^{\mu_a} \, ,
\ee
(the $\mu_a$'s should not be confused with the mass scale $\mu^4$ in front of 
$V$) and to introduce, as basis of independent scalars, the 4 symmetric 
polynomials
\be
\label{eq11}
\sigma_n \equiv \sum_a \mu_a^n \, .
\ee

With this notation, our first result is that the most general (densitized) 
potential can be written as
\be
\label{eq12}
\mu^4 \, {\cal V} (g_L , g_R) = \mu^4 (g_L \, g_R)^{\frac{1}{4}} \, V (\sigma_1 , 
\sigma_2 ,  \sigma_3 , \sigma_4) \, ,
\ee
where $V$ is an arbitrary function of the 4 $\sigma_a$'s.

\subsection{Universality classes}

In the same way as the various mathematical forms of the Landau free energy 
define universality classes of phase transitions, we can define {\it 
universality classes} of bigravity theories by considering as equivalent the 
functions $V(\sigma_a)$ leading to (essentially) the same multi-gravitational 
phenomenology.  As we shall see below and in \cite{DKP1,DKP2}, some of the 
important qualitative features of the function $V(\sigma_a)$ are: (i) its 
behaviour near $\sigma_a \rightarrow 0$, (ii) its behaviour when $\sigma_a 
\rightarrow \infty$, and (iii) the existence or non-existence of ``critical 
points'' where some derivatives $\partial V / \partial \sigma_a$ vanish.

As a first example of a universality class, we can define the class of $V$'s 
which reduce (in absence of cosmological constants in (\ref{eq6}), i.e. 
$\Lambda_L = \Lambda_R = 0$), in the linearized approximation, to the 
Pauli-Fierz mass term $\sim h_{\nu}^{\mu} \, h_{\mu}^{\nu} - h_{\mu}^{\mu} \, 
h_{\nu}^{\nu}$. The linearized approximation corresponds to the particular case 
where $g_{L \mu \nu}$ and $g_{R \mu \nu}$ are {\it both} near the {\it same} 
flat metric $\eta_{\mu\nu}$, i.e. $g_{L \mu \nu} = \eta_{\mu\nu} + 
h_{\mu\nu}^L$, $g_{R \mu \nu} = \eta_{\mu\nu} + h_{\mu\nu}^R$, with $h_L \ll 1$ 
and $h_R \ll 1$. In this limit the above object $H = g_L^{-1} \, g_R$ reads 
$H_{\nu}^{\mu} \simeq \delta_{\nu}^{\mu} + h_{R\nu}^{\mu} - h_{L\nu}^{\mu}$ 
(where the indices on $h_R$ and $h_L$ are raised by $\eta^{\mu\nu}$). It is then 
seen that the eigenvalues of $H$ are $\lambda_a \simeq 1 + \mu_a$, where $\mu_a 
\ll 1$ are the eigenvalues of $h_{R\nu}^{\mu} - h_{L\nu}^{\mu}$.
With the identification of the massive graviton mode $h_{\mu\nu}$ as $h_{\mu\nu} 
= h_{\mu\nu}^R - h_{\mu\nu}^L$ (see below), one then sees that the Pauli-Fierz 
mass term $\sim h_{\nu}^{\mu} \, h_{\mu}^{\nu} - h_{\mu}^{\mu} \, h_{\nu}^{\nu}$ 
is obtained if the function $V(\sigma_a)$ behaves (modulo a {\it positive} 
factor that can be absorbed in the mass scale $\mu^4$) as
\be
\label{eq13}
V(\sigma_a) \simeq \sigma_2 - \sigma_1^2 = \sum_a \mu_a^2 - \left( \sum_a \mu_a 
\right)^2 \, , \quad \hbox{when} \ \mu_a \rightarrow 0 \, .
\ee
The behaviour (\ref{eq13}) near $\mu_a \rightarrow 0$ defines the universality 
class of ``Pauli-Fierz-like'' bigravity. Note that one can imagine a case where 
the potential $V$ does not have quadratic terms when $\mu_a \rightarrow 0$. In 
the linearized approximation, one would see two massless gravitons, while the 
full theory would contain two interacting metric field $g_L , g_R$ (and only one 
common diffeomorphism invariance).

As a second example of the concept of universality class, we can define the 
class of potentials ${\cal V} (g_L , g_R)$ which are symmetric under the 
exchange $g_L \leftrightarrow g_R$. It is easily seen that under the exchange $L 
\leftrightarrow R$, the eigenvalues $\lambda_a$ get inverted $(\lambda_a 
\rightarrow \lambda_a^{-1})$ so that the logarithmic eigenvalues $\mu_a$ change 
sign: $\mu_a \rightarrow -\mu_a$. The class of exchange-symmetric potentials 
therefore corresponds to the class of functions $V(\mu_a)$ which are  even 
in the $\mu_a$'s. In terms of the $\sigma$'s this becomes $V (\sigma_1 , 
\sigma_2 , \sigma_3 , \sigma_4) = V (-\sigma_1 , \sigma_2 , -\sigma_3 , 
\sigma_4)$.

As a further example of universality class, we can consider the class of 
functions which depend only on the first two invariants $\sigma_1 \equiv \sum_a 
\mu_a$ and $\sigma_2 = \sum_a \mu_a^2 : V = V(\sigma_1 , \sigma_2)$. We shall 
see that this class appears naturally in brane models, and our (preliminary) 
investigations suggest that this class might be general enough to describe all 
the possible qualitative features of a general bigravity theory.

\subsection{Equations of motion}

The equations of motion derived from the bigravity action read
$$
2 \, M_L^2 \left(R_{\mu\nu} (g^L) - \frac{1}{2} \, g_{\mu\nu}^L \, R 
(g^L)\right) + \Lambda_L \, g_{\mu\nu}^L = t_{\mu\nu}^L + T_{\mu\nu}^L \, ,
$$
\be
\label{eq14}
2 \, M_R^2 \left(R_{\mu\nu} (g^R) - \frac{1}{2} \, g_{\mu\nu}^R \, R 
(g^R)\right) + \Lambda_R \, g_{\mu\nu}^R = t_{\mu\nu}^R + T_{\mu\nu}^R \, .
\ee
Here $T_L^{\mu\nu} \equiv 2 (-g_L)^{-1/2} \, \delta \, S_{\rm matter}^L / \delta 
\, g_{\mu\nu}^L$ denotes the stress-energy tensor of the matter on the left 
brane $(S_{\rm matter}^L = \int d^4 x \, \sqrt{-g_L} \, L (\Phi_L , g_L))$, 
while $t_L^{\mu\nu} \equiv g_L^{\mu\alpha} \, g_L^{\nu\beta} \, 
t_{\alpha\beta}^L$ denotes the effective stress-energy tensor (as seen on the 
left brane) associated to the coupling term $S_{\rm int} \equiv -\mu^4 \int d^4 
x (g_R \, g_L)^{1/4} \, V(g_L , g_R)$:
\be
\label{eq15}
t_L^{\mu\nu} \equiv \frac{2}{\sqrt{-g_L}} \, \frac{\delta \, S_{\rm int} (g_L , 
g_R)}{\delta \, g_{\mu\nu}^L} = - 2 \, \mu^4 \left( \frac{g_R}{g_L} 
\right)^{\frac{1}{4}} \left[ g^{\mu\nu}_L \, \frac{V}{4} + \frac{\partial \, 
V(g_L , g_R)}{\partial \, g_{\mu\nu}^L} \right] \, .
\ee
The corresponding expressions for the right brane are obtained by the exchange 
$L \leftrightarrow R$. For instance
\be
\label{eq16}
t_R^{\mu\nu} = -2 \, \mu^4 \left( \frac{g_L}{g_R} \right)^{\frac{1}{4}} \left[ 
g^{\mu\nu }_R \, \frac{V}{4} + \frac{\partial \, V(g_L , g_R)}{\partial \, 
g_{\mu\nu}^R} \right] \, .
\ee

The Bianchi identities $\left( D_L^{\nu} \left( R_{\mu\nu}^L - \frac{1}{2} \, 
R^L \, g_{\mu\nu}^L \right) \equiv 0 \right)$, and the conservation of the 
material energy tensor ($D_L^{\nu} \, T_{\mu\nu}^L = 0;$ when the {\it matter} 
equations of motion are satisfied) imply the constraints:
\be
\label{eq17}
D_L^{\nu} \, t_{\mu\nu}^L = 0 \quad \hbox{and} \quad D_R^{\nu} \, t_{\mu\nu}^R = 
0 \, .
\ee
Actually these two constraints are not independent because the invariance of 
$S_{\rm int}$ under the unbroken diagonal diffeomorphism group implies the {\it 
identity}
$$
\sqrt{-g_L}D_L^{\nu} \, t_{\mu\nu}^L +\sqrt{-g_R} D_R^{\nu} \, t_{\mu\nu}^R \equiv 0 \, .
$$
The explicit expressions of the derivative terms $\partial \, V / \partial \, 
g_{\mu\nu}^{L,R}$ in Eqs.~(\ref{eq15}), (\ref{eq16}) tends to be rather 
complicated. However, they acquire a simple form when written in the special 
frames with respect to which both $g_{\mu\nu}^L$ and $g_{\mu\nu}^R$ are 
diagonalized (such as in Eq.~(\ref{eq9})). The mixed components of 
$t_{\mu\nu}^L$ and $t_{\mu\nu}^R$ with respect to any such frame (which can 
differ from the particular $e_{\mu}^a$ of (\ref{eq9}) by arbitrary rescalings 
$e_{\mu}^a \rightarrow \zeta^a \, e_{\mu}^a$, because such rescalings leave $t_{La}^a$ 
and $t_{Ra}^a$ invariants) take the simple form: (no summation on the frame 
index $a$)
$$
t_{La}^a = -2 \, \mu^4 \, e^{\frac{1}{4} \sigma_1} \left( \frac{V}{4} - 
\frac{\partial \, V}{\partial \, \mu_a} \right) \, ,
$$
\be
\label{eq18}
t_{Ra}^a = -2 \, \mu^4 \, e^{-\frac{1}{4} \sigma_1} \left( \frac{V}{4} + 
\frac{\partial \, V}{\partial \, \mu_a} \right) \, ,
\ee
with vanishing of the off-diagonal components (we recall: $\sigma_1 \equiv 
\sum_b \mu_b$).

Here, we considered the scalar potential as a function of the $\mu_a$'s. If $V$ 
is given as a function of the $\sigma_n$'s, Eq.~(\ref{eq11}), the derivative 
entering Eqs.~(\ref{eq18}) takes the explicit:
\be
\label{eq19}
\frac{\partial \, V (\sigma_1 , \ldots , \sigma_4)}{\partial \, \mu_a} = 
\frac{\partial \, V}{\partial \, \sigma_1} + 2 \, \mu_a \, \frac{\partial \, 
V}{\partial \, \sigma_2} + 3 \, \mu_a^2 \, \frac{\partial \, V}{\partial \, 
\sigma_3} + 4 \, \mu_a^3 \, \frac{\partial \, V}{\partial \, \sigma_4} \, .
\ee
This explicit expression illustrates the third type ((iii)) of universality 
class mentioned above: If there exist ``critical points'' where 
$\partial_{\sigma_2} \, V = \partial_{\sigma_3} \, V = \partial_{\sigma_4} \, V 
= 0$ (without restriction on $\partial_{\sigma_1} \, V$), such points give rise 
to a $t_{L\nu}^{\mu}$ and a $t_{R\nu}^{\mu}$ with the local ``equation of 
state'' $t_{L0}^0 = t_{L1}^1 = t_{L2}^2 = t_{L3}^3$ (and similarly for $t_R$), 
i.e. such that $t_{\mu\nu}^L \propto g_{\mu\nu}^L$ and $t_{\mu\nu}^R \propto 
g_{\mu\nu}^R$. In some cases, such critical points can be ``fixed points'' and 
can give rise (in the ``vacuum case'', i.e. in absence of ``material'' 
$T_{\mu\nu}^{L,R}$) to bi-(A)dS solutions of the coupled field equations. Note 
in this respect that the ``perturbative limit'' $\mu_a = 0$ is a critical point 
in the sense that $\partial_{\mu_a} \, V = \partial_{\sigma_1} \, V$, 
independently of the value of $a$, so that $\mu_a = 0$ (i.e. $g_{\mu\nu}^L = 
g_{\mu\nu}^R$) can be a (perturbative) fixed point of the coupled vacuum 
equations, corresponding to a bi-(A)dS solution, if the corresponding (constant) curvature 
$\lambda$ ($R_{L\nu}^{\mu} = \lambda \, \delta_{\nu}^{\mu} = R_{R\nu}^{\mu}$) 
satisfies the two equations
$$
-2 \, \lambda \, M_L^2 + \Lambda_L = -2 \, \mu^4 \left[ \frac{V}{4} - 
\partial_{\sigma_1} \, V \right]_{\mu_a = 0} \, ,
$$
\be
\label{eq20}
-2 \, \lambda \, M_R^2 + \Lambda_R = -2 \, \mu^4 \left[ \frac{V}{4} + 
\partial_{\sigma_1} \, V \right]_{\mu_a = 0} \, .
\ee
In the ``Pauli-Fierz'' universality class the right-hand sides of 
Eqs.~(\ref{eq20}) vanish and one has the usual relation $\lambda = \Lambda_L / 
(2 \, M_L^2)$ (with the constraint $\Lambda_L / M_L^2 = \Lambda_R / M_R^2$). In 
more general classes the coupling between the two worlds can modify the usual 
link between $\lambda$ and $\Lambda_{L,R}$.

\subsection{Single ``massive graviton'' as a limiting case of bigravity}

Let us consider the formal limit $M_R \rightarrow \infty$ in the action 
(\ref{eq6}) (and the field equations (\ref{eq14})). In this limit the metric 
$g_R$ is (formally) frozen into some given ``background'' metric $G_{\mu\nu} : 
g_R \rightarrow G$, with $G_{\mu\nu}$ solution of $R_{\mu\nu} (G) = \lambda \, 
G_{\mu\nu}$, where $\lambda = \lim (\Lambda_R / (2 \, M_R^2))$  can be 
zero, or can be arranged to take any fixed real value. This leaves us with an 
action for a single dynamical metric $g \equiv g_L$ of the form
\be
\label{eq21}
S = \int d^4 x \sqrt{-g} \, (M_L^2 \, R(g) - \Lambda_L) - \mu^4 \int d^4 x \, (g 
\, G)^{\frac{1}{4}} \, V (G^{-1} \, g) \, .
\ee
If $V$ belongs to the Pauli-Fierz universality class, $g_{\mu\nu} = G_{\mu\nu}$ 
is a solution of the equations of motion (if $\lambda = \Lambda_L / (2 \, 
M_L^2))$, and the small excitations of $g_{\mu\nu}$ around $G_{\mu\nu}$ describe 
a ``massive graviton'' (propagating in an Einstein space). But the behaviour of 
the large excitations of $g_{\mu\nu}$ are described by the non-linear action 
(\ref{eq21}) instead of the usual quadratic Pauli-Fierz action.

The action (\ref{eq21}) is (formally) generally covariant: when $g$ is 
transformed as (\ref{eq4}), the frozen metric $G$ must also be transformed as 
$\delta \, G_{\mu\nu} = D_{\mu}^G \, \epsilon_{\nu} + D_{\nu}^G \, \epsilon_{\mu}$. These 
fluctuations of $G$ (which do not change the background curvature invariants) 
are playing the same role as the Goldstone degrees of freedom in the Higgs 
mechanism for gauge fields. In a recent paper \cite{Porrati:2001db} these 
Goldstone degrees of freedom were discussed for the single AdS$_4$ brane case, 
using an holographic description of five-dimensional gravity in terms of a 
four-dimensional CFT, and it was shown that there is indeed a vector field which 
provides extra components to the graviton.

\section{Specific Examples of Bigravity Effective \\ Lagrangians}

After having discussed general possible structural features of bigravity 
effective Lagrangians, we shall consider specific physical models in which such 
Lagrangians arise. We consider in turn: (i) brane models, (ii) Kaluza-Klein 
models, and (iii) non-commutative-type models. Beforehand let us note that the 
work in the seventies that first considered bigravity models did not have any 
underlying physical models from which they could derive some specific potentials 
$V(g_L , g_R)$. They made up some non-linear generalizations of the quadratic 
Pauli-Fierz mass term. For instance, they particularly considered the 
one-parameter family of models with
\be
\label{eq3.1}
V(g_L , g_R) \propto \left( \frac{g_R}{g_L} \right)^{a} (g_{\alpha \mu}^L 
\, g_{\beta\nu}^L - g_{\alpha\beta}^L \, g_{\mu\nu}^L) (g_R^{\alpha\beta} - 
g_L^{\alpha\beta}) (g_R^{\mu\nu} - g_L^{\mu\nu}) \, .
\ee

\subsection{Brane models}

Let us start by briefly recalling why (multi-)brane models naturally give rise 
to ``multigravity''. For more details the reader is advised to look at the 
original papers, and/or at reviews such as, \cite{Rubakov:2001kp}, 
\cite{Kogan:2001ub}. Before explaining how several worlds can be gravitationally 
``weakly coupled'', let us recall that the paradigmatic brane example of a 
separate (gravitationally decoupled) brane world is a Randall-Sundrum (RS) 
scenario, i.e. a flat 3-brane in AdS(5), with jump conditions on the brane 
(coming from an assumed $Z_2$ symmetry) able to ``localize'' the 5-dimensional 
graviton as a massless  excitation propagating (as a ``surface wave'') in the 
vicinity of the brane \cite{Gogberashvili:1999tb}, \cite{Randall:1999vf}. 
Putting the brane at the point $y=0$ (where $y$ is the ``fifth'', transversal 
coordinate, and where one requires the $Z_2$ symmetry $y \rightarrow -y$), the 
background 5-dimensional geometry is (see Fig.~2a)
\be
\label{eq3.2b}
ds^{2} = e^{-2 \sigma(y)} \eta_{\mu \nu} dx^{\mu} dx^{\nu} + dy^2 =
e^{-2 \sigma(z)} \left[\eta_{\mu \nu} dx^{\mu} dx^{\nu} + dz^2 \right] \, .
\ee
The warp factor behaves as $\sigma(y) \sim |y| \sim \ln( 1 + |z|)$.
The fluctuations near the background  metric are studied by writing:
\be
\label{eq3.3b}
ds^2 = e^{-2\sigma(z)} \left[ \left(\eta_{\mu\nu} + h_{\mu\nu} (x,z) \right) 
dx^\mu dx^\nu + dz^2 \right] \, .
\ee
The field $h_{\mu\nu}(x,z)$ is expanded in terms of the graviton and KK 
plane wave states : $ h_{\mu\nu} (x,z) = \sum_{n=0}^{\infty} \exp 
\left(\frac{3}{2} \, \sigma \right) h_{\mu\nu}^{(n)} (x) \Psi_n(z)$
where the $\exp \left(\frac{3}{2} \, \sigma \right)$ factor in the expansion
is necessary for the functions $\Psi^{(n)} (z)$ to obey an ordinary Schr\"{o}dinger 
equation:
\be
\label{eq3.4b}
\left\{ -\partial_z^2 + V(z) \right\} \Psi_{n} (z) = m_n^2 \Psi_{n} (y) \, .
\ee
Here  the potential $V(z) = (dA/dz)^2 - d^2 A/ dz^2$ where $A = 3 \sigma(z)/2$.
Qualitatively it is made up of an attractive $\delta$-function potential
plus a smoothing term (due to the AdS geometry) that gives the attractive 
potentials a ``volcano'' form. An interesting characteristic of this potential 
is that it gives rise to a (massless) normalizable zero mode 
\be
\label{eq3.5b}
\Psi_0(z) = \exp [-A(z)] = \exp[-\frac{3}{2} \, \sigma(z)]
\ee
One can show (see for details \cite{Csaki:2000fc} and references therein) that  
 the normalization factor $\int dz \Psi_0^2 = \int dz \exp[ -3 \sigma (z)]$ also
 relates  the fundamental five-dimensional mass scale $M_5$ to the  
four-dimensional Planck mass $M_p$, namely $M_p^2 = M_5^3 \int dz  \exp [-3 
\sigma (z)]$. 

\begin{figure}
\begin{center}
\epsfxsize=5in
\epsfbox{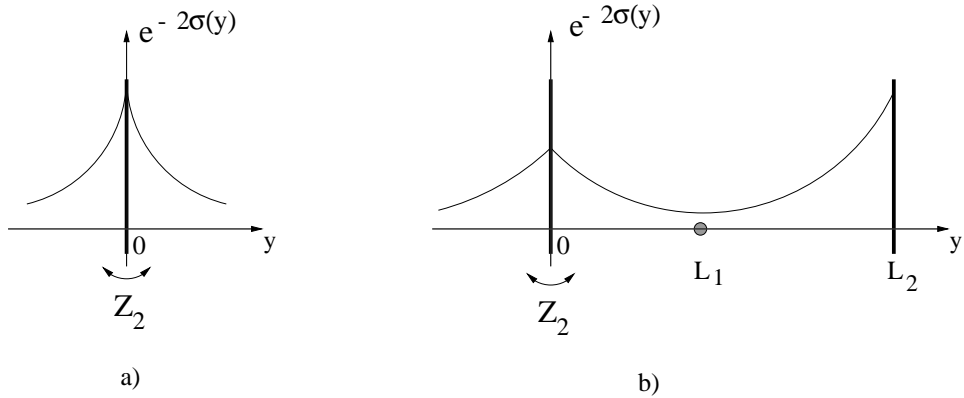}
\caption{ Warped metric for single flat brane (Fig.~2a) and bounce for a 
two-brane configuration (Fig.~2b). $L_2$ is the separation between branes and 
$L_1$ is the position of the bounce. If $L_1 \ll L_2$ the metric is mostly 
concentrated on a right brane and if $L_2-L_1 \ll L_1$ then it is concentrated  
on the left brane.}
\end{center}
\end{figure}

One can consider now multibrane configurations where the warped metric is 
a bounce as on Fig.~2b. By analysing the spectrum in this case one can easily 
see that in the  case of an infinitely large separation between the 
 branes  massless gravitons are  
localized on both of them.  But then,  according to basic properties of the 
Shr\"{o}dinger equation, when the separation is finite, 
 the degeneracy between the two massless modes is removed and one ends up with 
 one massless and one ultralight massive graviton. The prototype 
model of this class was the ``$+-+$ bigravity'' model \cite{Kogan:2000wc} with 
two positive tension flat branes ($''+''$ branes) separated at the bounce 
position $L_1$ by one intermediate negative tension flat brane ($''-''$ brane) in 
an $AdS_5$ bulk. The task of finding the KK spectrum reduces to a simple quantum 
mechanical problem. It is simple to see that the model (as every compact model) 
has a massless graviton that corresponds to the ground state of the system whose 
wave function follows the ``warp'' factor. Then it is easy to see  that (say, 
for simplicity, in the symmetric case $L_2 = 2L_1$) there
should be a state with wave function antisymmetric with respect to the minimum of 
the ``warp'' factor, whose mass splitting from the massless graviton will be 
very small compared to the masses  of the higher levels. Because the ``warp'' 
factor is exponential the difference  in  mass behaviour of the first and the 
rest of the KK states is also exponential. This allowed for the construction of  
a  linear ``bigravity'' model in which the remainder of the KK tower does not affect 
gravity beyond the millimetre bound. Soon after this model, other models were discussed.  
Some of them, for example the ``quasi-localized'' GRS model \cite{Gregory:2000jc}   
and a more general $+--+$ multigravity model \cite{Kogan:2000cv}, 
\cite{Kogan:2000xc} also used dynamical negative tension branes. In other 
models, like the  $++$ model with two $AdS_4$ branes \cite{Kogan:2001vb} (or the 
limiting case of one single $AdS_4$ brane  when the second one is moved to 
infinity \cite{Karch:2000ct}) or in a six-dimensional case \cite{Kogan:2001yr}, 
there are no negative tension branes and so no problems emerge with ghost-like
radion states. Models with moving branes in which one also can get 
``warped'' factors were discussed in \cite{Gorsky:2000aj}-\cite{Ellis:2000sx}.  
 Finally there is a  whole zoo of different models in which one can get modification 
of gravity at large scales.

To be specific, let us consider the $+-+$ model and let us now derive the fully 
non-linear bigravity action it gives rise to in the weak-coupling limit where we 
keep only the dominant terms in the exponentially small (``tunnelling'') 
coupling between the two positive tension branes. We are going to ignore the fact
 that  there is  a ghost-like radion field due to the existence of the negative tension
 brane  in this model \cite{Pilo:2000et}. Anyway we  freeze all
 dilaton and radion  degrees of freedom.

The action describing the full 5-dimensional configuration is (in units where 
the five-dimensional Planck mass is set to one)
\be
\label{eq3.2}
S = \int d^4 x \int dy \, \sqrt{-G} \left[ R(G) - V_b - \sum_i V_i \, \delta (y 
- y_i) \right] \, .
\ee
Here, $G$ denotes the 5-dimensional metric, $V_b$ the bulk cosmological 
constant, and $V_i$ the tensions of the branes (the index $i$ takes, in our 
case, three values corresponding to the three branes: e.g. $V_1 = V_L > 0$, $V_2 
< 0$ and $V_3 = V_R > 0$). We generalize the linear fluctuation ansatz 
 (\ref{eq3.3b}) by writing the 5-dimensional metric as ($\mu , \nu = 0,1,2,3$)
\be
\label{eq3.3}
ds_5^2 = a^2 (y) \, g_{\mu \nu} (x,y) \, dx^{\mu} \, dx^{\nu} + dy^2 \, ,
\ee
where $a^2 (y) = e^{-2 \sigma (y)}$ is the background warp factor. We assume 
here that the degrees of freedom associated to the fluctuations of: (i) the warp 
factor (``dilaton''), and (ii) the distance between the branes (``radions'') are 
all frozen. The detailed mechanism of how to do that  is not important for us now.
 For example,  one can  add extra terms in the 
action (\ref{eq3.2}) that give large enough mass terms to these fluctuations (say 
with submillimeter Compton wave length) following \cite{Goldberger:1999uk} (of course
 for those  radion fields which are ghost-like  one has to add  tachyonic mass terms).
 The fluctuations of the mixed components 
$g_{\mu y}$ can be consistently set to zero, because of the $Z_2$ symmetry 
requirement. Inserting the ansatz (\ref{eq3.3}) into the action (\ref{eq3.2}) 
yields, after integration by parts and use of the background equations of motion 
for the warp factor (which allow one to dispose of all terms containing 
$y$-gradients of $a(y)$)  the following action for $g_{\mu \nu} (x,y)$
\be
\label{eq3.4}
S = \int d^4 x \int dy \, \sqrt{-g(x,y)} \left\{ a^2 (y) \, R (g(x,y)) - 
\frac{1}{4} \, a^4 (y) \left[ {\rm tr} (g^{-1} \, \partial_y \, g)^2 - ({\rm tr} \, 
g^{-1} \, \partial_y \, g)^2 \right]\right\} \, .
\ee
Here, ${\rm tr} \, h^2 - ({\rm tr} \, h)^2 \equiv h_{\nu}^{\mu} \, h_{\mu}^{\nu}  - 
(h_{\mu}^{\mu})^2$ where $h_{\nu}^{\mu} \equiv (g^{-1} \, \partial_y \, 
g)_{\nu}^{\mu} = g^{\mu\sigma} \, \partial_y \, g_{\sigma \nu}$. Note that this 
exact (after freezing  the dilaton and the  radions) action for the nonlinear 
dynamics of $g_{\mu \nu} (x,y)$ is still 5-dimensional. Note also that all explicit 
coupling to the branes have disappeared (thanks to integration by parts). The 
crucial feature of (\ref{eq3.4}) for our discussion of an effective 4-dimensional 
bigravity action is the presence of the warp-factor dependent coefficients $a^2 
(y)$ and $a^4 (y)$. It is the fact that these factors are exponentially localized 
on the two positive-tension branes (as shown in Fig.~2b which plots $a^2 (y) = 
e^{-2\sigma(y)}$) which will allow for the derivation of an approximate 
4-dimensional action. Though Eq.~(\ref{eq3.4}) was derived from an explicit 
5-dimensional model, we expect that the general structure of Eq.~(\ref{eq3.4}), 
namely to have a curvature term with a weight function (here $a^2 (y) \, dy$) which 
localizes it on some branes, and a transverse-gradient term which also comes with a 
similarly localized weight function (here $a^4 (y) \, dy$), will hold in more 
general situations, like, for instance, the 6-dimensional model of 
\cite{Kogan:2001yr} (which is free from negative tension branes). 
Probably, in the latter model, if we assume that the 
excitations related to gradients in the sixth ``angular type'' direction $\theta$ 
are frozen (i.e. massive enough), we shall get an effective action of the type 
(\ref{eq3.4}) but, possibly, with weight factors which are somewhat modified (by 
the $y$-varying volume of the sixth circular dimension). To enhance the generality 
of our discussion, and cover such cases, we shall henceforth work with an action of 
the form (\ref{eq3.4}) but with the replacements $a^2 (y) \rightarrow a_2 (y)$, 
$a^4 (y) \rightarrow a_4 (y)$, where $a_2 (y)$ and $a_4 (y)$ are two (unrelated) 
``weight'' functions which are strongly localized around two branes. 
 The   generalization to the case of  $N$ branes is obvious. 
 The essential features of $a_2 (y)$ and 
$a_4 (y)$ that will be needed in the following is that they are both positive and 
that: (i) $a_2 (y)$ reaches maxima which are sharply localized on two branes, while 
(ii) $a_4^{-1} (y)$ reaches a sharp maximum somewhere between the two branes. The 
crucial point is to realize that these generic conditions imply the following 
specific $y$-dependence of $g_{\mu\nu} (x,y)$: as a function of $y$, $g(y)$ is 
nearly constant everywhere, except in a ``transition layer'', located around the 
{\it minimum} of $a_4 (y)$, where $g(y)$ has a fast variation with $y$. In other 
words $g(y)$ is a smoothed version of a Heaviside step function: $g(y) \simeq g_1 
\, \theta (y_* - y) + g_2 \, \theta (y-y_*)$ where $y_*$ is the location of the 
minimum of $a_4 (y)$ and where $g_1$ and $g_2$ are two {\it different} asymptotic 
values (which depend on $x^{\mu}$ when putting back everywhere the $x$-dependence). 
It is this transition-layer behaviour which allows us to derive an approximate 
4-dimensional action for $g_1 (x)$, $g_2 (x)$.

To understand intuitively this transition-layer behaviour we can assume that we 
normalize $a_4^{-1} (y)$ so that it takes the value $a_4^{-1} (y_*) = 1$ at its 
maximum and then decreases to very small values as $y$ gets away from $y_*$ (either 
way). Let us view the action (\ref{eq3.4}) (with $a^2 \rightarrow a_2$, $a^4 
\rightarrow a_4$) as a ``mechanical'' Lagrangian for the motion of the particle 
$g$, when thinking of $y$ as being ``time''. The ``kinetic'' terms are the last two 
terms quadratic in $y$-derivatives. We then view $a_4 (y)$ as the ``mass'' of the 
$g$-particle. This mass is of order unity around $y=y_*$, and then increases to 
very large values on both sides. In other words, the $g$-particle is extremely 
``heavy'' everywhere away from $y_*$, and becomes relatively ``light'' only around 
$y_*$, which makes it clear that $g(y)$ will ``move'' very little away from $y_*$, 
and that all ``$y$-motion'' will take place only around $y_*$. Another way of 
seeing that what is important is to have {\it separate} maxima in $a_2 (y)$ and 
$a_4^{-1} (y)$ would be to consider the $y$-Hamiltonian, (in terms of the 
$y$-momentum $\pi = \partial {\cal L} / \partial (\partial_y \, g)$) which is of 
the symbolic form: $a_2 \, R + a_4^{-1} \, \pi^2$. One can technically analyze 
the behaviour of $g(y)$ in the transition-layer by zooming on the exact solution of 
the only relevant part of the ``dynamics'' near $y = y_*$, namely the ``kinetic 
terms'' $\sim a_4 (\partial_y \, g)^2 \sim a_4^{-1} \, \pi^2$. Note that $a_2$ 
takes very small values around $y_*$, so that we can, in first approximation, 
neglect $a_2 \, R$ with respect to $a_4^{-1} \, \pi^2$. This can be done {\it 
exactly} by changing the ``time variable''. Indeed, in terms of the new ``time'' 
$t$, defined by
\be
\label{eq3.5}
dt \equiv a_4^{-1} (y) \, dy \, ,
\ee
the ``kinetic'' part of the action (\ref{eq3.4}) reads simply
\be
\label{eq3.6}
S_k = - \int d^4 x \, I \, ,
\ee
where
\be
\label{eq3.7}
I \equiv \frac{1}{4} \int dt \, \sqrt{- \det g(t)} \, \left[{\rm tr} (g^{-1} \, 
\dot g)^2 - ({\rm tr} \, g^{-1} \, \dot g)^2 \right] \, .
\ee
Here $\dot g \equiv \partial g / \partial t$, and we leave implicit the 
$x$-dependence of $g$. Actually, (\ref{eq3.6}) does not couple anymore $t$- and 
$x$-derivatives. Therefore we can solve the equations of motion derived from 
(\ref{eq3.6}) separately for each point $x$, i.e. it is enough to solve 
(\ref{eq3.7}) at each $x$. The action (\ref{eq3.7}) is still a very non-linear 
action for the $t$-dynamics of a $4 \times 4$ matrix $g_{\mu\nu} (t)$. However, it 
is exactly integrable. This is seen by exploiting the symmetries of $I$: (i) 
invariance under rigid SL(4) transformations of $g_{\mu\nu}$, and (ii) invariance 
under time translations. Note that we only have an SL(4) symmetry, and not a GL(4) 
one because of the presence of $\det g$. In other words, the action is invariant 
under $g_{\mu\nu} \rightarrow g'_{\mu\nu} = \Lambda_{\mu}^{\alpha} \, 
\Lambda_{\nu}^{\beta} \, g_{\alpha\beta}$ only when $\det \Lambda = 1$.
 The first  symmetry leads to the traceless mixed tensor constant of motion (in any spacetime 
dimension $D$)
\be
\label{eq3.8}
C_{\nu}^{\mu} = \pi_{\nu}^{\mu} - \frac{1}{D} \, \pi_{\sigma}^{\sigma} \, 
\delta_{\nu}^{\mu} \, .
\ee
Here $\pi_{\nu}^{\mu} \equiv g_{\nu\sigma} \, \pi^{\mu\sigma}$ where $\pi^{\mu\nu}$ 
is the ``momentum'' conjugate to $g_{\mu\nu}$, namely
\be
\label{eq3.9}
\pi^{\mu\nu} = \frac{\delta \, I}{\delta \, \dot{g}_{\mu\nu}} = \sqrt{-g} \, 
(K^{\mu\nu} - K \, g^{\mu\nu}) 
\ee
where $K^{\mu\nu} \equiv g^{\mu\alpha} \, g^{\nu\beta} \, K_{\alpha\beta}$ with 
$K_{\alpha\beta} = \frac{1}{2} \, \dot{g}_{\alpha\beta}$  being  the 
usual ``second fundamental form''.

The second symmetry leads to the constancy of the ``energy'':
\be
\label{eq3.10}
E = \frac{1}{4} \, \sqrt{-g} \, \left[{\rm tr} (g^{-1} \, \dot g)^2 - ({\rm tr} \, 
g^{-1} \, \dot g)^2 \right] = \sqrt{-g} \, (K_{\nu}^{\mu} \, K_{\mu}^{\nu} - K^2) \, 
.
\ee
Contrary to what happens in the well-known Kasner solutions, we are not restricted 
here to the ``zero-energy'' shell (because of the influence of the curvature term 
$a_2 \, R$ which changes the asymptotic behaviour of $g$ on both sides of the 
``transition layer'' that we are currently zooming into). This implies that the 
exact solution $g_{\mu\nu} (t)$ is different from, and more complicated than, a 
Kasner solution.

The exact solution is obtained by decomposing $g_{\mu\nu}$ in its determinant (or 
better $w \equiv \sqrt{-\det g}$) and its unimodular part, say $\gamma_{\mu\nu} / 
(-\det g)^{1/D}$. Eq.~(\ref{eq3.8}) simply says that $w \, \gamma^{-1} \, 
\dot\gamma$ is the constant matrix $2 \, C_{\nu}^{\mu}$. This is immediately 
integrated to the matrix equation
\be
\label{eq3.11}
\gamma (t) = \gamma (t_0) \exp \left( 2 \, C \int_{t_0}^t \frac{dt}{w(t)} \right) 
\, .
\ee
To complete the solution for $g(t)$ we need to know how its determinant $-\det g 
\equiv w^2$ depends on $t$. This is obtained by combining (\ref{eq3.8}) with 
(\ref{eq3.10}). This yields a first order differential equation for $w(t)$:
\be
\label{eq3.12}
\frac{D-1}{D} \, \dot{w}^2 = c^2 - E \, w \, ; \ c^2 \equiv {\rm tr} \, C^2 = 
C_{\nu}^{\mu} \, C_{\mu}^{\nu} \, .
\ee
In terms of the new parameter
\be
\label{eq3.13}
x \equiv \frac{1}{c} \, \left[ \sqrt{\frac{D}{D-1}} \, \frac{E}{2} \, t - B \right] 
\, ,
\ee
where $B$ is a constant of integration we get the solution
$$
w = \sqrt{-\det g} = \frac{c^2}{E} \, (1-x^2) \, ,
$$
\be
\label{eq3.14}
\frac{dt}{w} = \frac{2}{c} \, \sqrt{\frac{D-1}{D}} \, \frac{dx}{1-x^2} \, .
\ee
This allows one to express the matrix $g$ as an explicit function of $x$:
\be
\label{eq3.15}
g(x) = \left( \frac{1-x^2}{1 - x_0^2} \right)^{\frac{2}{D}} \, g(0) \exp \left( 4 
\sqrt{\frac{D-1}{D}} \, \widehat C \int_{x_0}^x \frac{dx}{1-x^2} \right) \, ,
\ee
where the matrix $\widehat C$ is $\widehat{C}_{\nu}^{\mu} \equiv c^{-1} \, 
C_{\nu}^{\mu}$, i.e. is normalized so that ${\rm tr} \, \widehat{C}^2 = 1$.

The above exact solution for $g(x)$, i.e. $g(t)$, using (\ref{eq3.13}), does not 
seem to involve any transition-layer behaviour. The transition-layer behaviour 
appears when we express $g$ in terms of the original transverse variable $y$ (which 
is the proper distance orthogonally to the branes). Indeed, when (qualitatively) 
integrating Eq.~(\ref{eq3.5}) to express $t$ as a function of $y$, the sharp 
maximum of $a_4^{-1} (y)$ around $y_*$ means that $t(y)$ behaves essentially as a 
(smoothed) step function $t(y) \simeq t_1 \, \theta (y_* - y) + t_2 \, \theta (y - 
y_*)$. Inserting this sharp-transition behaviour into the smooth solution 
$(g(x(t))$ (\ref{eq3.15}) then leads to the announced (smoothed) step-like 
behaviour of $g(y)$, with the bonus that we now have in hand the (rather 
complicated) precise manner in which $g(y)$ sharply (but smoothly) evolves in the 
transition region. It is interesting to make the link between the nonlinear 
transition of $g(x,y)$ between the two positive-tension branes (which is a smoothed 
version of $g(x,y) \simeq g_1 (x) \, \theta (y_* - y) + g_2 (x) \, \theta (y-y_*)$) 
and the result  of linearized fluctuations  which, as recalled above, is expressed as
$$
g_{\mu\nu} (x,y) = \eta_{\mu\nu} + h_{\mu\nu} (x,y) \quad \hbox{with} \quad 
h_{\mu\nu} (x,y) = \Sigma_n \exp \left(\frac{3}{2} \, \sigma \right) \, 
h_{\mu\nu}^{(n)} \, (x) \, \Psi_n (y) \, .
$$
One indeed finds, when looking at the explicit results for the various mode 
functions $\Psi_n (y)$ that the first two modes ($n=0,1$; corresponding to the 
massless mode, and the lightest mode) behave as $\Psi_0 (y) \equiv \exp \left( - 
\frac{3}{2} \, \sigma \right)$ and $\Psi_1 (y) \simeq \varepsilon (y-y_*) \exp 
\left( - \frac{3}{2} \, \sigma \right)$ where $\varepsilon (x) \equiv {\rm sign} 
(x)$. Keeping only the first two modes is then equivalent to considering metric 
fluctuations of the form $g_{\mu\nu} (x,y) \simeq \eta_{\mu\nu} + h_{\mu\nu}^{(0)} 
(x) + \varepsilon (y-y_*) \, h_{\mu\nu}^{(1)} (x)$, which is fully consistent with 
our result for the fully nonlinear metric $g(x,y)$ interpolating between a $g_1 
(x)$ and a $g_2 (x)$ through a transition layer. When going beyond the 
step-function approximation, one can also check that the nontrivial transition 
behaviour (\ref{eq3.15}) does also correspond (when linearized in $g_{\mu\nu} 
-\eta_{\mu\nu}$) to a zoom on the (large $k\ell$) limit of the first mode 
$e^{\frac{3}{2} \sigma} \, \Psi_1 (y)$ (considered as a smoothed version of 
$\varepsilon (y-y_*)$). Note also that the characteristic width of the transition 
layer is $\Delta y \sim k^{-1}$, where $k$ is the usual bulk curvature parameter 
defined such that the background solution has $(\partial_y \, \sigma)^2 = k^2$ 
(outside the branes), so that $\sigma (y) = k \, \vert y-y_1 \vert$ near brane $1$ 
(and $a^2 (y) = e^{-2\sigma} = e^{-2 k \vert y-y_1 \vert}$). There is a clean 
separation between the transition layer (around $y_*$ which is the location of the 
middle negative-tension brane) and the ``localization layers'' (around $y_1$ and 
$y_2$, i.e. the locations of the two positive-tension branes), when $k\ell \gg 1$, 
where $\ell$ denotes the smallest interbrane distance: $\ell = \min (L_1 , L_2 - 
L_1)$; see Fig.~2b. Because of the exponential dependence of the warp factors (and 
therefore of $a_2 (y)$ and $a_4 (y)$ in the $+-+$ model), even a moderately large 
value of $k\ell$ suffices to ensure that the above (nonlinear) transition-layer 
approximation is valid up to exponentially smaller corrections.

The exact, nonlinear transition-layer solution (\ref{eq3.15}) interpolates between 
a certain metric $g_1 (x) \equiv g(x,y_1)$ on the first brane, and another one $g_2 
(x) \equiv g(x,y_2)$ on the second brane. Instead of viewing the exact solution 
(\ref{eq3.15}) as the solution of a Cauchy problem (e.g. for given $g_1$ and 
$\dot{g}_1$), we should reexpress it as the solution of a ``Lagrange-Feynman'' 
problem, i.e. as the unique extremizing solution of the action (\ref{eq3.6}), 
(\ref{eq3.7}), for given ``initial'' and ``final'' values of $g(y)$: i.e. for given 
$g_1 (x) \equiv g(x,y_1)$ and $g_2 (x) \equiv g(x,y_2)$. We can also think of 
(\ref{eq3.7}) as defining a certain Riemannian metric in the space of metrics 
$g_{\mu\nu}$. We are then considering the ``geodesic'' connecting some given 
initial point $g_1$ to some given final point $g_2$. Let $g_{g_1 , g_2} (y)$ denote 
this unique (parametrized) geodesic. The analysis above then leads us to estimate 
that a good approximation (when $k\ell \gg 1$) to the effective action describing 
the dynamics of $g_1 (x)$ and $g_2 (x)$ is obtained by inserting the ``geodesic'' 
$g_{g_1 , g_2} (x,y)$ (computed for each point $x$) in the original full action 
(\ref{eq3.4}), so that
\be
\label{eq3.16}
S \, [g_1 , g_2] = \int d^4 x \, {\cal L} [g_1 , g_2] \, ,
\ee
where (suppressing the $x$-dependence to focus on the $y$-dependence)
\be
\label{eq3.17}
{\cal L} [g_1 , g_2] = 2\int_{y_1}^{y_2} dy \, a_2 (y) \, {\cal R} [g_{g_1 , g_2} 
(y)] - 2I [g_1 , g_2] \, ,
\ee
where ${\cal R} [g] \equiv \sqrt{-g} \, R(g)$, and where $I [g_1 , g_2]$ is the 
value of the ``geodesic'' action (\ref{eq3.7}) evaluated for the extremizing 
solution $g_{g_1 , g_2} (y)$ and integrated between $y_1$ and $y_2$. The factors
 $2$ in (\ref{eq3.17}) come from the fact that we are assuming periodicity over $y$ 
 varying between  $y_1$ and $2y_2-y_1$.
 Calculating $I [g_1 , g_2]$ from the exact solution 
(\ref{eq3.15}) is somewhat complicated. Let us only give the final result (which is 
simpler than the necessary intermediate steps):
\be
\label{eq3.18}
I [g_1 , g_2] = (g_1 \, g_2)^{\frac{1}{4}} \, V_b (g_1 , g_2) \, ,
\ee
where
\be
\label{eq3.19}
V_b (g_1 , g_2) = \frac{8}{T} \, \frac{D-1}{D} \, (\cosh \beta - \cosh \alpha)
\ee
with
\be
\label{eq3.20}
T = \int_{y_1}^{y_2} dt = \int_{y_1}^{y_2} a_4^{-1} (y) \, dy \, ,
\ee
\be
\label{eq3.21}
\beta = \frac{1}{4} \, \sqrt{\frac{D}{D-1}} \, \sqrt{\sigma_2 - 
\frac{\sigma_1^2}{D}} = \frac{1}{4} \, \sqrt{\frac{D}{D-1}} \, 
\sqrt{\widehat{\sigma}_2} \, ,
\ee
\be
\label{eq3.22}
\alpha = \frac{1}{4} \, \sigma_1 \, .
\ee
As above, $\sigma_1 \equiv \Sigma_a \, \mu_a$, $\sigma_2 \equiv \Sigma_a \, 
\mu_a^2$ where $\mu_a$ denote the logarithms of the eigenvalues of the matrix 
$g_1^{-1} \, g_2$, i.e. $\sigma_1 = {\rm tr} \, \ln \, g_1^{-1} \, g_2$ and 
$\sigma_2 = {\rm tr} \, (\ln \, g_1^{-1} \, g_2)^2$. The combination 
$\widehat{\sigma}_2$ is $\Sigma_a \, \widehat{\mu}_a^2$ where $\widehat{\mu}_a 
\equiv \mu_a - \sigma_1 / D$ denote the logarithms of the eigenvalues of the 
unimodular metric $\gamma_1^{-1} \, \gamma_2$, i.e. $\widehat{\sigma}_2 = {\rm tr} 
\, (\ln \, \gamma_1^{-1} \, \gamma_2)^2$. For added generality, we have left the dependence 
upon the brane (spacetime) dimension, though we have in mind here only $D=4$. The 
weak-coupling parameter appearing in front of the interaction term $I$ is the 
inverse of the total ``$t$-time'' $T = \int dt$ needed to interpolate between $g_1$ 
and $g_2$. We recall that, in the $+-+$ model, we have $a_4 = a^4$. An explicit 
computation then yields
\be
\label{eq3.23}
T = \frac{e^{4k\ell}}{2k} \, .
\ee
This exponentially large value (due to the exponentially small value of $a^4$ near 
the intermediate brane) corresponds to the expected exponentially small coupling 
between the two metrics on the positive-tension branes.

To get an explicit bigravity action, one still needs to evaluate the first 
contribution in the Lagrangian (\ref{eq3.17}). Neglecting exponentially small 
fractional contribution it is clear (in view of the localized behaviour of $a_2 
(y)$ and of the near $y$-constancy of $g_{g_1 , g_2} (y)$ outside of the 
transition-layer) that this contribution is well approximated by replacing $g_{g_1 
, g_2} (y)$ by its (relevant) boundary value $g_1$ or $g_2$. Finally, the full 
brane-derived bigravity effective Lagrangian density (in units where the 
coefficient of ${\cal R}$ in the higher-dimensional theory is set to one) is
$$
{\cal L} (g_1 , g_2) = A_1 \, {\cal R} (g_1) + A_2 \, {\cal R} (g_2) - 2I (g_1 , 
g_2) \, ,
$$
where ${\cal R} (g) \equiv \sqrt{-g} \, R(g)$, where (see Fig.~2b; we assume that $y$ 
varies over a full period $[-L_2, +L_2]$)
\be
\label{eq3.24}
A_1 = \int_{-L_1}^{L_1} dy \, a_2 (y) \, , \quad A_2 = \int_{L_1}^{2L_2 - L_1} dy 
\, a_2 (y) \, ,
\ee
and where the ``potential'' term $I(g_1 , g_2)$ is given by Eqs.~(\ref{eq3.18}), 
(\ref{eq3.19}) above.

It is easily checked that the potential (\ref{eq3.19}) has the Pauli-Fierz limiting 
behaviour (\ref{eq13}) in the limit $\mu_a \rightarrow 0$. One can then compute the 
corresponding Pauli-Fierz mass. One finds (in the symmetric case $L_2 = 2 \, L_1 = 
2 \, \ell$, for simplicity) $m_{\rm PF}^2 = 4 \, T^{-1} \, A_1^{-1}$. The explicit 
value, in the $+-+$ model, of $A_1$ is $A_1 \simeq 1/k$, so that we get $m_{\rm PF} 
= 2 \, \sqrt 2 \, e^{-2k\ell} \, k$, in agreement with the direct analysis of 
linearized fluctuations \cite{Kogan:2000wc}. In the Appendix  we further compare the 
nonlinear bigravity action to the linearized bigravity results already derived in 
the literature. In particular we check that they are fully consistent, even in the 
asymetric case $L_2 \ne 2 \, L_1$.

A full justification of the effective action (\ref{eq3.17}) can, in principle, be 
obtained by explicitly considering the effect of corrections to our approximation 
$g(x,y) \simeq g_{g_1 , g_2} (x,y)$. For instance, we can write $g(y) \equiv g_{g_1 
, g_2} (y) + \xi (y)$ where the correction $\xi (y)$ vanishes, by definition, when 
$y=y_1$ and $y=y_2$. We can then expand $\xi (y) = \Sigma_n \, \xi_n \sin 2\pi \, n 
\, \tau (y)$ where $\tau \equiv t/T$ varies (by definition) in the interval 
$(0,1)$. [The condensed notation $\xi_n$ denotes some $\xi_{n\mu\nu} (x)$]. An 
analysis of the full action (expanded quadratically in the $\xi$'s), containing not 
only the ``light fields'' $g_1 (x)$, $g_2 (x)$, but the tower of ``heavy fields'' 
$\xi_n (x)$, shows that the mass of the heavy fields scale like $m_{\xi} \sim 
e^{-k\ell} \, k$, which is exponentially heavier (by a factor $e^{+k\ell}$) than 
the Pauli-Fierz mass scale. This confirms that the nonlinear bigravity action 
(\ref{eq3.17}) is a good effective description when one considers configurations 
$g_1 (x)$, $g_2 (x)$ where the relevant gradients are small compared to $m_{\xi}$.

\subsection{Kaluza-Klein Models}

As said in the Introduction, and sketched in Fig.~1, one expects generic 
Kaluza-Klein models to give rise to ``regular spectra'' containing no gap allowing 
one to separate a finite number of light gravitons from an infinite tower of heavy 
ones. We wish, however, to emphasize the existence of a class of KK models where 
such a gap can exist.

By KK model, we mean a higher-dimensional background geometry which decomposes as a 
direct (unwarped) product, $ds_{\rm tot}^2 = ds_D^2 + d\sigma^2$ where $ds_D^2 = 
g_{\mu \nu}^{(0)} (x) \, dx^{\mu} \, dx^{\nu}$ (with, e.g., $g_{\mu \nu}^{(0)} (x) 
= \eta_{\mu\nu}$) and $d\sigma^2 = \gamma_{ab} (y) \, dy^a \, dy^b$. When 
decomposing the fluctuations of the higher-dimensional metric $g_{MN} (x,y) = 
(g_{\mu\nu} , g_{\mu a} , g_{ab})$ into representations of the symmetry group of 
$g_{\mu\nu}^{(0)}$ (say $g_{\mu\nu}^{(0)} = \eta_{\mu\nu}$) one generally expects 
the squared mass spectrum of tensor (spin 2) fluctuations $\delta \, g_{\mu\nu}$ to 
be given by the spectrum of the {\it scalar} Laplacian on the (compact) internal 
manifold, say $\Gamma$, with metric $d\sigma^2 = \gamma_{ab} (y) \, dy^a \, dy^b$. 
Let $\lambda_n \geq 0$, with $n=0,1,2,\ldots$ denote the latter spectrum, i.e. 
$\gamma^{-1/2} \, \partial_a (\gamma^{1/2} \, \gamma^{ab} \, \partial_b) \, \phi_n 
(y) = -\lambda_n \, \phi_n (y)$. There is always a zero-mode, $\phi_0 (y) = {\rm 
const.}$, corresponding to $\lambda_0 = 0$, i.e. to a massless graviton. The 
question of the existence of a hierarchy allowing one to consider, for instance, an 
effective theory containing only the massless graviton and a superlight one, is 
then equivalent to requiring that the first eigenvalue $\lambda_1$ (or group of 
eigenvalues) be parametrically smaller than higher eigenvalues. It is interesting 
to note that there are general mathematical theorems which guarantee that such a 
hierarchy cannot occur if the compact metric $\gamma$ is Ricci-flat (or 
Ricci-positive). Indeed, if we consider, for simplicity, the Ricci-flat case, there 
are theorems (see \cite{berard}, \cite{ledoux}) saying that 
there exist {\it universal} positive constants $a_n (d)$, $b_n (d)$ (which depend 
only on the dimension $d$ of the compact manifold $\Gamma$) such that $a_n (d) \, 
\delta^{-2} < \lambda_n < b_n (d) \, \delta^{-2}$ for all $n \geq 1$, where 
$\delta$ denotes the (metric) ``diameter'' of $\Gamma$. However, we wish to 
emphasize that, if one does not constraint the sign of the Ricci tensor, nothing 
prevents the occurrence of a spectral hierarchy. We conjecture that the generic 
situation where such a spectral hierarchy (between a finite group of abnormally 
small eigenvalues and the rest) occurs is a ``near pinching'' situation, i.e. the case 
where the manifold $\Gamma$ is on the verge of getting split into two (or more) 
separate manifolds (of the same dimension $d$ as $\Gamma$), as is illustrated in 
Fig.3a.
\begin{figure}
\begin{center}
\epsfxsize=5in
\epsfbox{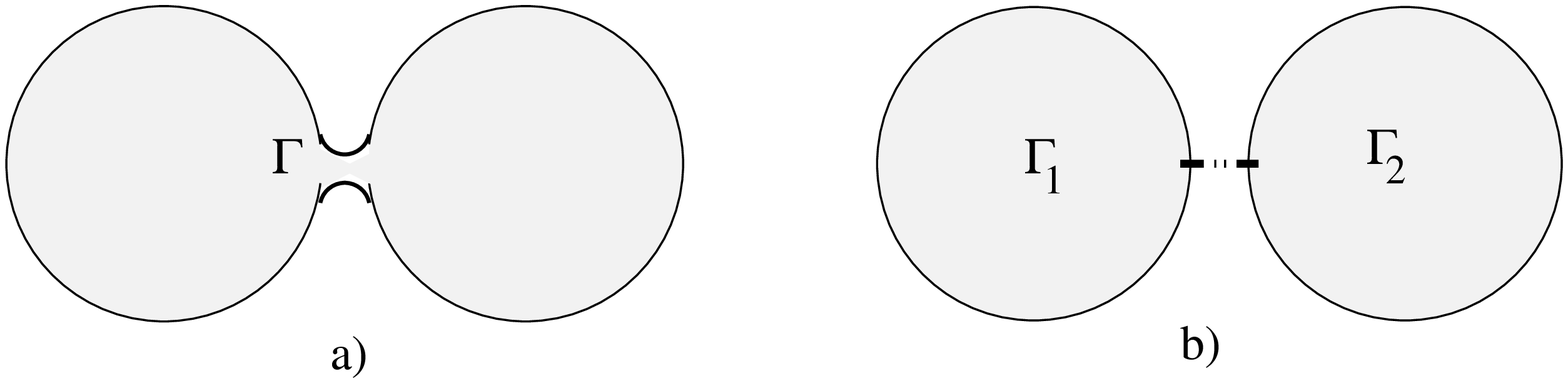}
\caption{Manifold $\Gamma$ (Fig. 3a) is on the  verge of 
splitting into two classically disconnected
 manifolds $\Gamma_1$ and $\Gamma_2$ (Fig. 3b). 
These two manifolds may be connected at the quantum  level.} 
\end{center}
\end{figure}

 We have confirmed by some toy-model calculations that the near-pinching case 
(if the connecting ``tube'' between, say, two manifolds is not too long) does indeed 
lead to a spectral hierarchy. Let us also mention that a general theorem of Cheeger 
(see \cite{berger}) can be viewed as a (moral) confirmation of our conjecture. 
Indeed, this theorem says that a {\it lower bound} of the first eigenvalue is 
$\lambda_1 \geq h^2 / 4$ where Cheeger's constant $h$ is defined as the lower bound of 
the ratio $\vert S \vert / {\rm inf} (\vert \Gamma_1 \vert , \vert \Gamma_2 \vert)$ 
when $S$ runs over all closed submanifolds of $\Gamma$ (of dimension $d-1$) which 
partition $\Gamma$ into two open manifolds $\Gamma_1 , \Gamma_2$, with common boundary 
$\partial \Gamma_1 = \partial \Gamma_2 = S$. We use the notation $\vert \Gamma \vert$ 
to denote the (riemannian) volume of $\Gamma$. Note that $h$ has the dimension of an 
inverse length, and that ``pinching'' does indeed correspond to the case where $h 
\rightarrow 0$.

Physically, we can view the very light mode arising in a nearly pinched configuration 
$\Gamma_1 \cup \{{\rm tube}\} \cup \Gamma_2$ as coming from the effect of a weak 
coupling between two ``resonators'' (or quantum mechanical systems) having regular 
spectra $\lambda_n^{(1)}$, $\lambda_n^{(2)}$. Before coupling, the ground state is 
degenerate, $\lambda_0^{(1)} = \lambda_0^{(2)} = 0$. Weak coupling is generally 
expected to split this degeneracy into a doublet. As $\lambda_0 = 0$ is always an exact 
eigenvalue of the combined system (corresponding to $\phi_0 (y) =1$), this mechanism 
always leads to a small $\lambda_1$ (going to zero with the coupling). Note that the 
eigenmode corresponding to $\lambda_1$ is approximately equal to $\phi_1 (y) \simeq 
\varepsilon (y)$ where $\varepsilon (y)$ is $+1$ over $\Gamma_1$, and $-1$ over 
$\Gamma_2$.

We are aware of the fact that weakly coupled string theory suggests compactification 
on Ricci-flat manifolds (which exclude a spectral hierarchy). However, we think that 
string theory might still, in certain circumstances, allow for a spectral hierarchy: 
either because of $\alpha'$-corrections to Einstein equations (which lead one away 
from the Ricci-flat case), or (more speculatively) because of conceivable quantum 
tunnelling effect between two (separate, but ``near'') Ricci-flat manifolds 
$\Gamma_1$, $\Gamma_2$. Pictorially, such a tunnelling situation is the limit of 
Fig. 3b where the link between $\Gamma_1$ and $\Gamma_2$ is classically broken. The 
exponentially small coupling associated to such a tunnelling situation would naturally 
induce an exponentially small $\lambda_1$, and thereby a bigravity coupling scale 
$\mu$ exponentially smaller than the string scale. 

\subsection{Bigravity and Connes' non-commutative geometry}

Within his general non-commutative geometry programme \cite{connes}, Connes introduced 
the model of a two-sheeted space $X$, made from the product of a continuous space $Y$ 
by a discrete ``two-point space'' $\{ a,b \}$ (or $Z_2$) : $X = Y \times Z_2$. Though 
the algebra ${\cal A}$ of ``functions on $X$'' (defined as the algebra of pairs of 
functions viewed as diagonal matrices ${\rm diag} (f_a (y) , f_b(y))$ with $y \in Y$) 
is commutative, the bimodule of 1-forms on such a space is not commutative 
\cite{connes}. Generalizations of this model (also based on the product of a continuum 
by a discrete space) were used in \cite{Connes:1990qp} to give a geometrical 
explanation of the structure of the Standard Model. In particular, it was found that 
the VEV of the Higgs field is related to the (non-commutative) ``distance'' between 
the two sheets. The {\it metric} aspect of such a two-sheeted space was developed 
along different lines by several authors \cite{Chamseddine:1992yx}, 
\cite{Chamseddine:ms}, \cite{Chamseddine:1995xh}, \cite{Chamseddine:1996zu}. For 
instance, Ref.~\cite{Chamseddine:1992yx} introduced (non-commutative) analogues of the 
Riemannian metric, curvature tensor and scalar curvature, which enabled them to 
introduce a generalized Einstein-Hilbert action. This generalized Einstein-Hilbert 
action was found to contain (besides the standard integral of the scalar curvature of 
$Y$) a minimally coupled massless scalar field $\sigma$ related to the ``distance'' 
between the two sheets by $d \propto e^{-\sigma (y)}$. An alternative approach to 
studying gravitational effects within general non-commutative spaces has been proposed 
in \cite{Chamseddine:1996zu}. We shall follow this approach 
which is based on a general ``spectral action principle''. In its simplest form, this 
principle is proposing to take as bare bosonic (Euclidean) action for any 
non-commutative model $X$ the trace of the heat kernel associated with the square of 
the (non-commutative) Dirac operator $D_X$ of $X$:
\be
\label{eq3.25}
I = {\rm Tr} \, \exp \, (-t \, D_X^2) \, .
\ee
Here $t \equiv m_0^{-2}$ introduces a cut-off, roughly equivalent to keeping only 
frequencies smaller than $m_0$. The cut-off-dependent Euclidean action (\ref{eq3.25}) 
is viewed (\`a la Wilson) as the bare action at the mass scale $\sim m_0$.

It seems that all previous works interested in the metric aspect of a two-sheeted 
space $X = Y \times Z_2$ have restricted themselves (either for simplicity, or because 
of some constraints \cite{Chamseddine:1995xh}) to the case where the metric is the 
same on the two sheets. By contrast, we focus here on the case where the two metrics 
are different, say $g_{\mu\nu}^L$ and $g_{\mu\nu}^R$, and the aim of this subsection 
is to compute the ``potential'' ${\cal V} (g_L , g_R)$ implied by the spectral action 
(\ref{eq3.25}). Following Connes (see p.~569 of the English edition of his book 
\cite{connes}) we define a Dirac operator on a bi-Riemannian space $X = Y \times Z_2$ 
as
\be
\label{eq3.26}
D_X = \left(\begin{array}{ccc}
D\!\!\!\!/\,_L  & \, & \gamma_5 \, m \\
\, & \, & \, \\ 
\gamma_5 \, m & \, & D\!\!\!\!/\,_R \\
\end{array}\right) \, .
\ee

This operator acts on bi-spinors $\left( {\psi_L \atop \psi_R} \right)$ living on $Y 
\times Z_2$. Our conventions are that the (Euclidean) gamma matrices are hermitian, as 
well as $\gamma_5$ (which satisfies $\gamma_5^2 = 1$ and which anticommutes with the 
gamma matrices and therefore with the separate Dirac operators $D\!\!\!\!/\,_L$ and 
$D\!\!\!\!/\,_R$). The explicit form of the (hermitian) Dirac operators on each sheet 
is
\be
\label{eq3.27}
D\!\!\!\!/\,_{L,R} = i \, \gamma^{\mu}_{L,R} \left(\partial_{\mu} + 
\Omega_{\mu} (L,R) \right), \quad \{ \gamma^{\mu}_{I}, \gamma^{\nu}_{I} \} = 2 \, 
g^{\mu\nu}_{I}, \quad I = L,R \, .
\ee

The explicit form of the spin connections $\Omega_{\mu}$ will not be important for our 
calculations. On the other hand, the explicit form of the gamma matrices will be 
crucial. They read $\gamma_L^{\mu} = \gamma^a \, E_{aL}^{\mu}$, $\gamma_R^{\mu} = 
\gamma^a \, E_{aR}^{\mu}$ where $\{ \gamma^a , \gamma^b \} = 2 \, \delta^{ab}$ is a 
standard set of (space-independent) gamma matrices and where $E_{aL}^{\mu} (x)$, 
$E_{aR}^{\mu} (x)$ (where $x \in Y$) are vierbeins corresponding to the two 
 positive definite metrics $g_{\mu\nu}^L$, $g_{\mu\nu}^R$ given on the abstract 
 manifold $Y : g_L^{\mu\nu} (x) = \delta^{ab} \, E_{aL}^{\mu} \, E_{bL}^{\nu}$, etc. 
 Note that the structure  (\ref{eq3.26}) assumes that we are given not only an 
 identification map between 
corresponding points of the two sheets (here gauge-fixed by the identification of the 
two underlying abstract manifolds and the use of only one coordinate system $x^{\mu}$ 
to describe the metrics on the two sheets), but also a one-to-one map between the spin 
structures, and in particular between any choice of vierbein. In other words to any 
$E_{aL}^{\mu}$ must correspond a unique $E_{aR}^{\mu}$ so that an arbitrary, local 
SO(4) rotation of $E_{aL}^{\mu}$ corresponds to the same rotation of $E_{aR}^{\mu}$. 
It is most natural to use as map $E_{aL}^{\mu} \rightarrow E_{aR}^{\mu}$ the canonical 
map defined in \cite{jpb}. This map can be defined by requiring that it reduces to 
simple rescalings $E_{aR}^{\mu} = e^{-\mu_a / 2} \, E_{aL}^{\mu}$ when considering 
bi-orthogonal frames (as in Eq.~(\ref{eq9}) above). For simplicity, the quantity $m$ 
in Eq.~(\ref{eq3.26}) (which ``connects'' the two sheets) will be taken to be a 
constant real scalar. More generally, it could also  be a matrix when considering multiplets 
 of fermions and could be space-dependent. We 
shall see that $m$ is connected with the coupling scale $\mu$ in Eq.~(\ref{eq6}). It 
might be interesting to consider generalized models where $m$ (and therefore $\mu$) is 
linked to a fluctuating scalar $m (x) \propto e^{-\sigma (x)}$ as in 
\cite{Chamseddine:1992yx}.

The square of the Dirac operator (\ref{eq3.26}) is easily obtained as
\be
\label{eq3.28}
D_X^2 = \left(\begin{array}{ccc}
D\!\!\!\!/\,_L^2 + m^2 & \, &m \, \gamma_5( D\!\!\!\!/\,_R - D\!\!\!\!/\,_L ) \\
\, & \, & \, \\ 
m \, \gamma_5 (D\!\!\!\!/\,_L - D\!\!\!\!/\,_R) & \, &D\!\!\!\!/\,_R ^2 + m^2 \\
\end{array}\right)
\ee

The heat kernel expansion of (\ref{eq3.25}) is a series in increasing powers of 
$t=m_0^{-2}$ which starts at order $t^{-2} = m_0^4$. At this leading order the action 
$I$ leads to two bare cosmological constant terms $m_0^4 / (4 \pi^2) \int d^4 x 
(\sqrt{g_L} + \sqrt{g_R})$. At the next to leading order, ${\cal O} (t^{-1}) = {\cal 
O} (m_0^2)$, one gets two separate Einstein actions $m_0^2 / (48 \pi^2) \int d^4 x 
(-\sqrt{g_L} \, R_L - \sqrt{g_R} \, R_R)$ (with negative signs, as is appropriate for 
an Euclidean action $I_E$ which is essentially $I_E = - I_{\rm Minkowski}$ with a 
positive signature metric) as well as a ``potential'' term $+ \int d^4 x \, {\cal V} 
(g_L , g_R)$ proportional to $m_0^2 \, m^2$. 
In view of its $m_0^2 \, m^2$ scaling, the potential ${\cal V}$ contains no 
derivatives of $g_L$ or $g_R$. It can therefore be evaluated by considering the case 
of constant metrics $g_{\mu\nu}^L$, $g_{\mu\nu}^R$. We can then neglect the spin 
connections in (\ref{eq3.27}) and go to the momentum representation $(i \, 
\partial_{\mu} \rightarrow k_{\mu})$ to set
\be
\label{eq3.29}
D_X^2 = \left(\begin{array}{ccc}
k\!\!\!\!/\,_L^2 + m^2 & \, &m \, \gamma_5 (k\!\!\!\!/\,_R - k\!\!\!\!/\,_L ) \\
\, & \, & \, \\ 
m \, \gamma_5 (k\!\!\!\!/\,_L - k\!\!\!\!/\,_R) & \, &k\!\!\!\!/\,_R ^2 + m^2 \\
\end{array}\right) \, .
\ee
Here $k\!\!\!\!/\,_L \equiv \gamma_L^{\mu} \, k_{\mu}$, $k\!\!\!\!/\,_R \equiv 
\gamma_R^{\mu} \, k_{\mu}$. Using the explicit vierbein expressions of 
$\gamma_L^{\mu}$ and $\gamma_R^{\mu}$ we can rewrite these as $k\!\!\!\!/\,_L \equiv 
\gamma^a \, k_a^L$, $k\!\!\!\!/\,_R \equiv \gamma^a \, k_a^R$, where
\be
\label{eq3.30}
k_a^L \equiv k_{\mu} \, E_{aL}^{\mu} \ , \quad k_a^R \equiv k_{\mu} \, E_{aR}^{\mu} \, 
.
\ee
In terms of these two different vectors (that live in a local Euclidean space common 
to the tangent spaces of the two sheets), one easily finds that the eigenvalues of 
$D_X^2$ are
\be
\label{eq3.31}
\lambda_{\pm} = \frac{\mbox{\boldmath$k$}_L^2 + \mbox{\boldmath$k$}_R^2}{2} + m^2 \pm 
\sqrt{\frac{(\mbox{\boldmath$k$}_L^2 - \mbox{\boldmath$k$}_R^2)^2}{4} + m^2 
(\mbox{\boldmath$k$}_L - \mbox{\boldmath$k$}_R)^2} \, .
\ee
Here, all squares are evaluated with the flat Euclidean metric $\delta^{ab}$ 
appropriate to the local Euclidean space where both $k_a^L$ and $k_a^R$ live. In the 
limit $m^2 \rightarrow 0$ (appropriate to the heat kernel expansion) the eigenvalues 
(\ref{eq3.31}) read (we henceforth suppress the boldfacing of the Euclidean vectors 
$k^L$ and $k^R$)
$$
\lambda_+ = k_L^2 + m^2 + m^2 \, \frac{(k_L - k_R)^2}{k_L^2 - k_R^2} + {\cal O} (m^4) 
\, ,
$$
\be
\label{eq3.32}
\lambda_- = k_R^2 + m^2 - m^2 \, \frac{(k_L - k_R)^2}{k_L^2 - k_R^2} + {\cal O} (m^4) 
\, .
\ee
The heat kernel action reads 
\be
\label{eq3.33}
I = {\rm Tr} \, \exp \, (-t \, D_X^2) = 4 \int d^4 x \int \frac{d^4 k}{(2\pi)^4} \ 
[e^{-t \lambda_+} + e^{-t \lambda_-}],
\ee
where the 4 comes from the trace in spinor space and where $d^4 k$  is the fourfold integral 
 over the {\it covariant} components $k_{\mu}$. Expanding (\ref{eq3.33}) in powers of $m^2$
 leads to the mixing term $+ \, 4 \, 
m_0^2 \, m^2 \int d^4 x$ $(V_1 + V_2)$ where
\be
\label{eq3.34}
V_1 = -t^2 \int \frac{d^4 k}{(2\pi)^4} \ \frac{e^{-t k_L^2} - e^{-t k_R^2}}{k_L^2 - 
k_R^2} \, (k_L - k_R)^2 \, ,
\ee
\be
\label{eq3.35}
V_2 = -t^2 \int \frac{d^4 k}{(2\pi)^4} \, (e^{-t k_L^2} + e^{-t k_R^2}) \, .
\ee
After our factorization of $m_0^2 = t^{-1}$ in front of $V_1$ and $V_2$, the 
expressions (\ref{eq3.34}), (\ref{eq3.35}) are easily seen to be $t$-independent. We 
can then evaluate them by setting, say, $t=1$ in them. Noting that $k_L^2 = 
\delta^{ab} \, k_a^L \, k_b^L = g_L^{\mu \nu} \, k_{\mu} \, k_{\nu}$, etc., $V_2$ is 
easily evaluated:
\be
\label{eq3.36}
V_2 = -c_4 ((\det g_L^{\mu\nu})^{-\frac{1}{2}} + (\det g_R^{\mu\nu})^{-\frac{1}{2}}) = 
-c_4 (\sqrt{g_L} + \sqrt{g_R})
\ee
where $c_4 \equiv (16 \pi^2)^{-1}$ and where $g_L \equiv \det g_{\mu\nu}^L = (\det 
g_L^{\mu\nu})^{-1}$. The potential $V_1$ is much more tricky. However, it can be 
nicely expressed by introducing a Schwinger-type parameter $\alpha$ (varying between 0 
and 1) and by using the identity $(e^{-a} - e^{-b}) / (a-b) \equiv - \int_0^1 d\alpha 
\exp [-((1-\alpha) a + \alpha b)]$. This naturally leads to the introduction of a 
one-parameter family of metrics $g(\alpha)$ interpolating between $g_L$ and $g_R$ 
(reached, respectively, when $\alpha = 0$ and $\alpha = 1$). More precisely we define
\be
\label{eq3.37}
g^{\mu\nu} (\alpha) \equiv (1-\alpha) \, g_L^{\mu\nu} + \alpha \, g_R^{\mu\nu} \, .
\ee
[Note that the ``line'' connecting $g_L$ to $g_R$ is ``straight'' when expressed in 
terms of contravariant metrics (which naturally appear in the squared Dirac operator 
$D^2 = - g^{\mu\nu} \, \partial_{\mu\nu} = + \, g^{\mu\nu} \, k_{\mu} \, k_{\nu}$), but 
will become ``curved'' when expressed in terms of the covariant components $g_{\mu\nu} 
(\alpha)$.] In terms of the definition (\ref{eq3.37}) (and the associated $g_{\mu 
\sigma} (\alpha) \, g^{\sigma \nu} (\alpha) = \delta_{\mu}^{\nu}$, $g(\alpha) \equiv 
\det g_{\mu\nu} (\alpha) \equiv (\det g^{\mu\nu} (\alpha))^{-1}$) we find that $V_1$ 
can be written as
\be
\label{eq3.38}
V_1 = + \frac{1}{2} \, c_4 \, \Delta E_a^{\mu} \, \Delta E_a^{\nu} \int_0^1 d\alpha \, 
\sqrt{g(\alpha)} \, g_{\mu\nu} (\alpha) \, ,
\ee
where $\Delta E_a^{\mu} \equiv E_{aR}^{\mu} - E_{aL}^{\mu}$. Finally, the ${\cal O} 
(m^2)$ piece of the Euclidean action (i.e. the ``potential'', remembering that ${\cal 
L}_{\rm Eucl.} = - {\cal L}_{\rm Mink.}$) predicted by the non-commutative approach to 
two-sheeted spaces reads $+ \int d^4 x \, {\cal V}$ with
\be
\label{eq3.39}
{\cal V} = \frac{m_0^2 \, m^2}{4\pi^2} \left[ \frac{1}{2} \, \Delta E_a^{\mu} \, 
\Delta E_a^{\nu} \int_0^1 d\alpha \, \sqrt{g(\alpha)} \, g_{\mu\nu} (\alpha) - 
\sqrt{g_L} - \sqrt{g_R} \right] \, .
\ee

The explicit evaluation of the $\alpha$-integral in (\ref{eq3.39}) can be reduced to 
(incomplete) elliptic integrals. In fact, it can be reduced to the evaluation of the 
single integral
\be
\label{eq3.40}
I (g_L , g_R) \equiv \int_0^1 d\alpha \, \sqrt{g(\alpha)} = \int_0^1 
\frac{d\alpha}{\sqrt{\det ((1-\alpha) \, g_L^{\mu\nu} + \alpha \, g_R^{\mu\nu})}}
\ee
by using the identity
\be
\label{eq3.41}
\left( \frac{\partial}{\partial \, g_L^{\mu\nu}} + \frac{\partial}{\partial \, 
g_R^{\mu}} \right) I (g_L , g_R) = - \frac{1}{2} \int_0^1 d\alpha \, \sqrt{g(\alpha)} 
\, g_{\mu\nu} (\alpha) \, .
\ee
When considering a bi-orthogonal frame, say with $E_{aL}^{\mu} = {\rm diag} 
(e^{-\mu_a^L / 2})$, $E_{aR}^{\mu} = {\rm diag} (e^{-\mu_a^R / 2})$ (so that 
$g_{\mu\nu}^L = {\rm diag} (e^{+\mu_a^L})$, $g_{\mu\nu}^R = {\rm diag} (e^{+\mu_a^R})$ 
and $g_L^{-1} \, g_R = {\rm diag} (e^{\mu_a})$ with $\mu_a = \mu_a^R - \mu_a^L$) the 
integral (\ref{eq3.40}) is a rather simple elliptic integral of the first kind which, 
in principle, can be explicitly expressed in terms of the eigenvalues $\lambda_a^{L,R} 
\equiv e^{\mu_a^{L,R}}$. Of more direct interest for us is the discussion of the 
``weak-excitation'' limit of ${\cal V}$, i.e. the limit $\mu_a = \mu_a^R - \mu_a^L 
\rightarrow 0$, i.e. $(g_L^{-1} \, g_R)_{\nu}^{\mu} = \delta_{\nu}^{\mu} + 
h_{\nu}^{\mu}$ with $h_{\nu}^{\mu} \rightarrow 0$. In this limit we find that ${\cal 
V}$ behaves as (with $\sigma_1 = \Sigma \, \mu_a$, $\sigma_2 = \Sigma \, \mu_a^2$ as 
above)
\begin{eqnarray}
\label{eq3.42}
{\cal V} &\simeq &\frac{m_0^2 \, m^2}{4\pi^2} \, (g_L \, g_R)^{\frac{1}{4}} \left( -2 
+ \frac{\sigma_2}{8} - \frac{\sigma_1^2}{16} \right) \nonumber \\
&= &\frac{m_0^2 \, m^2}{4\pi^2} \, (g_L \, g_R)^{\frac{1}{4}} \left( -2 + \frac{1}{8} 
\, h_{\mu}^{\nu} \, h_{\nu}^{\mu} - \frac{1}{16} \, (h_{\mu}^{\nu})^2 \right) \, .
\end{eqnarray}

Besides a negative $m^2$-dependent, contribution to the cosmological constant (which 
has anyway bare contributions ${\cal O} (m_0^4)$), we see that we do not get a 
Pauli-Fierz-type mass term for weak excitations away from $g_{\mu\nu}^L = 
g_{\mu\nu}^R$. We get instead (remembering that the (bare) Planck mass is $M_L^2 = 
M_R^2 = m_0^2 / (48 \pi^2)$) a mass term proportional to $m^2 \left[ h_{\mu}^{\nu} \, 
h_{\nu}^{\mu} - \frac{1}{2} \, (h_{\mu}^{\nu})^2 \right]$. Such a mass term contains a 
scalar ghost, but has the virtue (contrary to the Pauli-Fierz one $\sim m^2 \, 
[h_{\mu}^{\nu} \, h_{\nu}^{\mu} - (h_{\mu}^{\mu})^2]$) of exhibiting  excellent continuity 
properties of the limit $m^2 \rightarrow 0$ for all processes linked to the generation 
of gravitational fields by sources (see, e.g., Appendix~C of \cite{Boulware:1972my} 
where it is easily seen that $\alpha = 1/2$ leads to an Einstein-like propagator 
$h_{\mu\nu} \, T^{\mu\nu} \propto T^{\mu\nu} (\Box - m^2)^{-1} \, \left( T_{\mu\nu} - 
\frac{1}{2} \, \eta_{\mu\nu} \, T \right)$).

\section{Phenomenology of Multigravity}

\subsection{Bigravity, and bicosmology, versus massive gravity}

There is quite a sizable (and somewhat confusing) literature about the ``problems'' 
raised by having either ``massive gravity'' (i.e. a kind of finite-range version of 
Einstein's theory), or a ``massive graviton'' in addition to Einstein's massless one. 
We leave to a future publication a detailed discussion of such issues, but wish to 
emphasize the fact that the change of paradigm, brought by focusing on a fully 
nonlinear bigravity theory, drastically modifies, in our opinion, the way one should 
view the traditional ``problems'' of massive gravity (in both senses recalled in the 
sentence above). One of the basic points is that many of the ``problematic'' issues 
(such as, unboundedness of the energy, singularity of the infinite-range limit) simply 
loose their meaning in a general bigravity setting. Indeed, these problematic issues 
make sense only for states (in some theories) which are, at least asymptotically, 
close to some trivial, {\it Poincar\'e invariant} background. We think that, even when 
considering formally ``small'' excitations above a trivial background state 
$g_{\mu\nu}^L (x) = g_{\mu\nu}^R (x) = \eta_{\mu\nu}$, the exact bigravity 
configurations will generically develop into full-blown ``bi-cosmological'' 
configurations with fields that grow so much (in time and/or in space) so as to be 
outside the usually considered domain of bi-asymptotically flat configurations 
containing localized excitations. Note that most of the results concerning the 
``discontinuity'' of the $m^2 \rightarrow 0$ limit \cite{vanDam:1970vg}, 
\cite{Zakharov}, \cite{Boulware:1972my} implicitly (or explicitly) assumed such a 
framework of asymptotically decaying perturbations of a (minimum energy) Poincar\'e 
invariant background. We think that, if one relaxes this asymptotic restriction, there 
exists a sector of bigravity theories which exhibits ``physical continuity'' for small 
$m^2$, at the cost of cosmological behaviour on large scales. Note that such a claim, 
while being consistent with the works \cite{Higuchi:1989gz,Kogan:2000uy,Porrati:2000cp}
 which  found continuity of massive graviton 
interactions in maximally symmetric ((A)dS) cosmological backgrounds, is somewhat 
different from the claim of \cite{vainshtein},\cite{deffayet}. Indeed, the latter 
claim seems to insist on a framework (and a language, like that of propagators, 
coupling and scattering states) which preassumes the restriction to localized 
excitations of a Poincar\'e-invariant vacuum, i.e. that the 
 metrics under consideration are   asymptotically flat.  
Leaving to a future publication a detailed discussion of the ``discontinuity'' issue, 
we shall content ourselves here to sketch the general dynamical structure \`a la 
Arnowitt-Deser-Misner (ADM) \cite{ADM} of bigravity theories.

\subsection{ADM analysis of bigravity theories}

We consider a general bigravity action (\ref{eq6}). Let us decompose the two spacetime 
metrics $^4 g_{\mu\nu}^L$, $^4 g_{\mu\nu}^R$ into the two lapses $N_L$, $N_R$ $(N_L 
\equiv (- \, ^4g_L^{00})^{-\frac{1}{2}})$, the two shift vectors $N_L^i$, $N_R^i$ 
$(^4g_{ij}^L \, N_L^j \equiv \, ^4g_{0i}^L)$ and the two spatial metrics $g_{ij}^L$, 
$g_{ij}^R$ $(g_{ij}^L \equiv \, ^4 g_{ij}^L)$. We have
$$
ds_L^2 = - N_L^2 \, dt^2 + g_{ij}^L (dx^i + N_L^i \, dt) (dx^j + N_L^j \, dt) \, ,
$$
\be
\label{eq4.1}
ds_R^2 = - N_R^2 \, dt^2 + g_{ij}^R (dx^i + N_R^i \, dt) (dx^j + N_R^j \, dt) \, .
\ee
After integration by parts, each separate ``Left'' or ``Right'' pieces of the action 
(\ref{eq6}) reads (say for the Left piece)
\be
\label{eq4.2}
I_L = \int dt \int d^3 x \, (\pi_L^{ij} \, \dot{g}_{ij}^L + \Pi_L \, \dot{\Phi}_L - 
N_L \, H_L - N_L^i \, H_i^L)
\ee
where $\pi_L^{ij}$ is the Left gravitational momentum density, $\Pi_L$ is a (generic) 
matter momentum density and where the left super-Hamiltonian, and super-momentum 
densities have the structure
\be
\label{eq4.2b}
H_L = \frac{1}{M_L^2} \ \frac{1}{\sqrt{g_L}} \left( \pi_L^{ij} \, \pi_{ij}^L - 
\frac{1}{2} \, \pi_L^2 \right) - M_L^2 \, \sqrt{g_L} \, R_L + H_L^{\rm matter} \, ,
\ee
\be
\label{eq4.3}
H_i^L = - 2 \ D_j^L \, \pi_{Li}^j + H_i^{L \, {\rm matter}} \, .
\ee
Let us now consider the interaction term $-(^4 g_L \, ^4 g_R)^{\frac{1}{4}} \, V (^4 
g_L^{-1} \, ^4 g_R)$. Using the fact that the local scalar $V$ must be (in particular) 
invariant under transformations of the type $dt' = \lambda \, dt$, $dx'^i = 
\Lambda_j^i (dx^j + v^j \, dt)$ one finds that it can only depend on the lapses and shifts 
 through the combinations   $N_R / N_L$ and 
 $(N_R^i - N_L^i) / \sqrt{N_L \, N_R}$. Let 
us then replace the 8 variables $N_L$, $N_L^i$, $N_R$, $N_R^i$ by the combinations
\be
\label{eq4.4}
\overline N \equiv \sqrt{N_R \, N_L} \, , \ n \equiv \sqrt{\frac{N_R}{N_L}} \, , \ 
\overline{N}^i \equiv \frac{1}{2} \, (N_R^i + N_L^i) \, , \ n^i \equiv \frac{N_R^i - 
N_L^i}{2 \, \sqrt{N_R \, N_L}} \, .
\ee
With these definitions it is found that the total action reads
\be
\label{eq4.5}
I =  \int dt \int d^3 x \, (\pi_L^{ij} \, \dot{g}_{ij}^L + \pi_R^{ij} \, \dot{g}_{ij}^R 
+ \Pi_L \, \dot{\Phi}_L + \Pi_R \, \dot{\Phi}_R - {\cal H})
\ee
where the total Hamiltonian density reads (here and below $g_{ij}^L$, $g_{ij}^R$ are 
the spatial metrics and $g_L \equiv \det g_{ij}^L$, $g_R \equiv \det g_{ij}^R$)
\be
\label{eq4.6}
{\cal H}(\overline N , \overline{N}^i , n , n^i , g_L , \pi_L , g_R , \pi_R , \Phi_L , \Pi_L 
, \Phi_R , \Pi_R) = \overline N \, \overline H + \overline{N}^i \, \overline{H}_i
\ee
where
\be
\label{eq4.7}
\overline H (n , n^i ,g , \pi , \Phi , \Pi) = n^{-1} \, H_L + n \, H_R - n^i \, H_i^L 
+ n^i \, H_i^R + (g_L \, g_R)^{\frac{1}{4}} \, V (n , n^i , g_L , g_R) \, ,
\ee
\be
\label{eq4.8}
\overline{H}_i (g , \pi , \Phi , \Pi) = H_i^L + H_i^R \, .
\ee

The crucial point for the present discussion is the separation of the 8 lapse and 
shift variables into two sets: (i) the four ``average'' lapse and shifts $\overline 
N$, $\overline{N}^i$, which are true Lagrange multipliers  appearing only {\it 
linearly} in the action, and (ii) the four ``relative'' lapse and shifts $n$, $n^i$ 
which enter algebraically in the action (no kinetic terms) but in a {\it non linear} 
manner. The four average lapse and shifts give rise to four constraints, which are 
linked to the symmetry of the action under common diffeomorphisms:
\be
\label{eq4.9}
\overline H = 0 \ , \qquad \overline{H}_i = 0 \, .
\ee
$\overline N$, $\overline{N}^i$ are gauge variables which can be gauged away (e.g. to 
$\overline N = 1$, $\overline{N}_i = 0$). The four (first-class) constraints 
(\ref{eq4.9}) can be used, together with the field equations (which involve $\overline 
N$ and $\overline{N}^i$) to eliminate four degrees of freedom (i.e. eight functions of 
positions and momenta). By contrast, the four relative lapse and shifts $n$, $n^i$ are 
not (undeterminable) gauge variables but are dynamical variables which are 
instantaneously determinable in terms of the other variables $(g,\pi ,\Phi ,\Pi)$ by 
their (algebraic) equations of motion:
\be
\label{eq4.10}
\frac{\partial \, \overline H}{\partial \, n} = 0 \ , \qquad \frac{\partial \, 
\overline H}{\partial \, n^i} = 0 \, .
\ee
This result generalizes the findings of \cite{Boulware:1972my} which studied the case of
 ``massive gravity'', i.e. (\ref{eq21}).
We must assume here that the potential $V$ has a ``good'' dependence on $n$ and $n^i$ 
which allow for an (essential) unique solution of Eqs.~(\ref{eq4.10}) for a generic (or 
at least an open) domain of free dynamical data $g, \pi , \Phi , \Pi$. We think that 
the only (covariant) situation where $n$ and $n_i$ combine with $\overline N$, 
$\overline{N}^i$ to generate more (gauge-related) Lagrange multipliers is the case 
where ${\cal V}$ is linear in $N_L$ and $N_R$, which must then correspond (by 
covariance) to ${\cal V} = c_L \, N_L \, \sqrt{g_L} + c_R \, N_R \, \sqrt{g_R} = c_L 
(-\det \, ^4 g_{\mu \nu}^L)^{1/2} + c_R (-\det \, ^4 g_{\mu\nu}^R)^{1/2}$. 
For instance, we can 
think that $V(n,n_i)$ contains terms quadratic in $n^i$ (as already follows from a 
Pauli-Fierz mass term), and behaves, for both large (respectively, small) $n$ as $+ \, 
a \, n^2$ (respectively $+ \, b \, n^{-2}$). 
 Note also  that if we define the new scalar potential $V^{\rm new}$ by factoring 
$(-\det \, ^4 g_{\mu \nu}^L)^{1/2} + (-\det \, ^4 g_{\mu\nu}^R)^{1/2}$ instead of
 $(^4 g^L \,  ^4 g^R)^{1/4}$ from ${\cal V}$, i.e.  ${\cal V} =  
((-\det \, ^4 g_{\mu \nu}^L)^{1/2} + (-\det \, ^4 g_{\mu\nu}^R)^{1/2})V^{\rm new}$, the last 
term in Eq.~(\ref{eq4.7}) will become  $(n^{-1}\sqrt{g_L} + n\sqrt{g_R})V^{\rm new}(n,n_i,
g_L,g_R)$. It is then enough to require  that $V^{new}(n)$ grows in any manner (even 
 logarithmically) towards $+\infty$, as $n \rightarrow +\infty$ or $n \rightarrow 0$.
Such conditions ensure the existence of 
(possibly non unique) solutions of the equations of motion of $n$ and $n^i$, 
Eq.~(\ref{eq4.10}).

We can then use (\ref{eq4.10}) to eliminate $n$ and $n^i$ (by replacing them by their 
expression in terms of the other dynamical data). It is then easily seen that the 
reduced Hamiltonian $\overline{H}_{\rm red} (g,\pi ,\Phi ,\Pi)$ obtained by inserting 
these expressions into (\ref{eq4.7}) defines (together with (\ref{eq4.8})) a dynamical 
system for the variables $g_{ij}^L$, $\pi_L^{ij}$, $g_{ij}^R$, $\pi_R^{ij}$, $\Phi_L$, 
$\Pi_L$, $\Phi_R$, $\Pi_R$ (submitted to the four first-class constraints (\ref{eq4.9}) 
coming from the Lagrange multipliers $\overline N$, $\overline{N}^i$). For instance, if 
we consider the matter-free system, we end up with the $6+6$ degrees of freedom linked 
to $g_{ij}^L$ and $g_{ij}^R$, from which must be subtracted 4 degrees of freedom killed 
by the first-class constraints (\ref{eq4.9}). This leaves us with 8 degrees of freedom. 
As in the analysis of (nonlinear) massive gravity in \cite{Boulware:1972my}, which 
concluded to the presence of 6 degrees of freedom (instead of the expected 5 of a 
Pauli-Fierz linear graviton), we have here $2+6=8$ (where the 2 can be formally thought 
of as corresponding  to an  Einstein (massless) graviton, and the 6 to a ``massive graviton'').

Two of the potential defects of the supplementary tensor degree of freedom ($1=6-5$) 
are, according to \cite{Boulware:1972my}: (i) the unboundedness of the total 
``energy'', and (ii) experimental difficulties (e.g. with light scattering by the Sun), 
even if a suitable mass term can be found for which the $m^2 \rightarrow 0$ limit 
exists. Our point of view concerning (i) is to argue that the notion of energy is not 
defined when considering (as we argue must be done) non-asymptotically flat metrics, 
with cosmological-type behaviour at infinity. Alternatively, we can dismiss  the problem 
of spatial boundary conditions by considering spatially compact manifolds (e.g. with 
toroidal topology). For such a situation, the dynamics associated to 
(\ref{eq4.5})--(\ref{eq4.10}) should entail a well-defined (classical) evolution system 
for $g_L , \pi_L , g_R , \pi_R , \ldots$ The ill-defined issue of ``unbounded energy'' 
is then transformed in a well-posed dynamical question: do Hamilton's equations of 
motion quickly lead to a catastrophic evolution towards some singular state?, or do 
they admit many solutions which evolve rather quietly on times scales comparable to the 
age of the universe? (which is the only stability property which is really required by 
experimental data). This question will be discussed in detail in \cite{DKP1}. Let us 
only mention here the result that there does exist, for suitable potentials ${\cal V}$, 
many solutions which can quietly evolve on Hubble time scales or more.

\subsection{Phenomenology and a new form of dark energy}

Using the dynamical, and cosmological like, viewpoint expressed in the previous 
subsection, let us now briefly discuss why we think that bigravity is not only 
compatible with existing gravitational data, but might also furnish a natural 
explanation of the recently observed cosmic acceleration. Let us first argue that there 
exist large classes of bigravity data $g_L , \pi_L , g_R , \pi_R , \ldots$ which can 
adequately represent the universe as we see it at the present moment. For definiteness, 
we assume that we ``live on the right brane'' (when viewed in brane language), i.e. 
that the matter around as is made of $\Phi_R$-type matter only. Let us start by 
considering an instantaneous ``Einstein'' model of our universe, i.e. an exact solution 
$g_R , \pi_R , \Phi_R , \Pi_R$ of the constraints $H^R = 0 = H_i^R$. Let us complete 
this configuration by a random ``Einstein'' model  of the (shadow) left universe, i.e. 
a solution $g_L , \pi_L , \Phi_L , \Pi_L$ of $H^L = 0 = H_i^L$. Taken together, these 
two configurations ``nearly'' satisfy the bigravity constraints (\ref{eq4.9}) and 
(\ref{eq4.10}). More precisely, (\ref{eq4.9}) is satisfied modulo a term proportional 
to $V$, while (\ref{eq4.10}) is satisfied modulo terms proportional to $\partial V / 
\partial n$ and $\partial V / \partial n^i$. Let us assume that all dimensionless 
variables ($n$, $n^i$ and $g_L^{-1} \, g_R$) are of order unity, and that $V \propto 
\mu^4$ is at most comparable to the average cosmological energy density $V \sim 
10^{-29} \, g \, {\rm cm}^{-3}$ (i.e. $\mu \sim 10^{-3}$ eV) (in right units, say). 
Instead of viewing Eqs.~(\ref{eq4.10}) as equations for determining $n$ and $n^i$, we 
can pick rather arbitrary (initial) values of $n$ and $n^i$ (or order unity) and 
slightly deform the Einstein data $g_R , \pi_R , \ldots , g_L , \pi_L , \ldots$ to 
compensate for the small violation of the usual Einstein constraints brought by the 
terms proportional to $V$, $\partial V / \partial n$ and $\partial V / \partial n^i$. 
It is intuitively clear that there are many ways of doing so, i.e. of constructing 
exact bigravity initial data $g_R , \pi_R , \ldots , g_L , \pi_L , \ldots$ which 
exactly satisfy (\ref{eq4.9}) and (\ref{eq4.10}) for arbitrarily given $n$ and $n^i$. 
Locally, say around our Galaxy, the new, deformed data $g_R , \pi_R , \ldots$ can be 
constructed so as to be experimentally indistinguishable from a pure Einstein model 
(after all, we are simply modifying the stress-energy tensor in the Galaxy by less than 
$10^{-29} \, g \, {\rm cm}^{-3}$, which is many orders of magnitude smaller than the 
average density in the Galaxy). If the dependence of $V$ on $n$ and $n^i$ is adequate 
the equations (\ref{eq4.10}) will continue to admit a solution $(n,n^i)$ during the 
future evolution of the other dynamical variables. In fact, as (under a general 
assumption made in Section~2 above) we know  
one exact (but physically trivial) solution of the 
full bigravity evolution equations, namely  $g_{\mu\nu}^L (t,x) = g_{\mu\nu}^R (t,x)$, i.e. 
$g_L = g_R$, $\pi_L = \pi_R$, $\Phi_L = \Phi_R , \ldots$ with $n=1$, $n^i = 0$, we 
expect (by mathematical continuity) that there will be classes of bigravity solutions 
where, during a long time, $g_L \simeq g_R$, $\pi_L \simeq \pi_R , \ldots$ with $n 
\simeq 1$, $n^i \simeq 0$. The crucial question is whether one can solve 
Eqs.~(\ref{eq4.10}) for a long time (without catastrophe) for more general data where 
$g_L^{-1} \, g_R = {\cal O} (1)$ and $n = {\cal O} (1) = n^i$. This question will be 
addressed in \cite{DKP1} for cosmological-type solutions and in \cite{DKP2}  for 
solar-system-type solutions. Note that this is here that the potential 
``discontinuity'' problems linked to the $m^2 \rightarrow 0$ (or $\mu \rightarrow 0$) 
limit show up because the potential $V(n,n^i)$ is proportional to $\mu^4$, so that, 
when solving for $n$ and $n^i$ Eqs.~(\ref{eq4.10}), $\mu^4$ will tend to appear in a 
denominator and might cause the solution $n,n^i$ to take parametrically large values, 
proportional to some negative power of $\mu$ (depending on the behaviour of, say, 
$V(n)$ as $n \rightarrow +\infty$ or $n \rightarrow 0$).

Assuming, for the time being, the continuous existence of regular bigravity solutions, 
evolved from some data $g_L , \pi_L , g_R , \pi_R , n , n^i , \ldots$ we can finish by 
mentioning some of the pleasing phenomenological aspects of bigravity. First, bigravity 
exactly satisfies the equivalence principle, because each type of matter (say $\Phi_R$ 
whithin ``our universe'') is universally coupled to the corresponding metric, say 
$g_{\mu\nu}^R$. Second, (as just discussed) there are classes of bigravity solutions 
which differ from standard Einstein ones only by the presence ``on the right-hand 
side'' of Einstein's equations of  numerically very small additional terms (say 
$t_{\mu\nu}^R \sim t_{\mu\nu}^L \sim  10^{-29} \, g \, {\rm cm}^3$ in the covariant 
form (\ref{eq14})), which {\it locally} modify $g_{\mu\nu}^R$ and $\Phi^R$ (and 
$g_{\mu\nu}^L$, $\Phi^L$) only in a numerically very small way (though they might {\it 
globally} forbid the stable existence of asymptotically flat models). These solutions 
will be fully compatible with all local (or quasi-local) experimental tests of 
relativistic gravity: such as solar-system tests and binary-pulsar tests. Third, if 
$\mu$ indeed happens to be of the order of $10^{-3}$ eV, and if $g_L^{-1} \, g_R$ is of 
order unity, bigravity will only lead to experimentally significant deviations from 
Einstein's gravity on cosmological scales. Moreover, if, seen from our universe 
$g_{\mu\nu}^R$, we view $g_{\mu\nu}^L$ as an ``external field'', or, more precisely, if 
we (approximately) view the ``difference'' between the two metrics $g_L^{-1} \, g_R$ as 
a given (time varying) tensor ``condensate'' of order unity, the potential term ${\cal V} = 
\sqrt{g_R} (g_R^{-1} \, g_L)^{\frac{1}{4}} \, V (g_L^{-1} \, g_R)$ can be approximately 
viewed as a time-varying ``vacuum energy'' term (of order $\mu^4$), i.e. as a kind of 
``dark energy''. It is tempting to assume that this new form of dark energy (which might 
 be called ``tensor quintessence'') can explain 
the observed cosmic acceleration. It might also be used in primordial cosmological 
scenarios, possibly when using the idea mentioned above that $\mu$ could be an evolving 
field.  See \cite{DKP1} for a study of this new form of dark energy, and its 
phenomenological differences with quintessence models based on evolving scalar (rather 
than tensor) condensates.

\section{Conclusions}

 In this paper we suggested a new  paradigm concerning ``massive gravity'' and ``large
 scale modification of gravity''.  Considering the fully nonlinear  bigravity action 
suggests to change viewpoint: instead of the theory with massless and massive graviton(s) we
 had in linearized approximation, we are dealing with  several interacting metrics.
 We introduced the  concept of universality class which we formulated using bigravity
 (two interacting metrics) as an example. Different approaches 
(brane, KK, non-commutative geometry) naturally lead to different universality classes 
 for the fully nonlinear bigravity action.
 Another important new  suggestion  is that almost all solutions must  now be  of 
 the  non-asymptotically flat (cosmological) type.

This new formulation can change the standard problematic of the  $m^2 \rightarrow 0$ 
discontinuity. We showed the existence of classes of solutions that are compatible with 
``our universe''. However, we do not claim to have proven that general solutions of 
bigravity are phenomenologically acceptable. The two main problems of massive gravity 
(ghost, potential blow up of some field variables when $m^2 \rightarrow 0$) must still 
be examined in detail. The  important  problem is to find  the  matching 
to the local sources of the field so that the full metric is free of singularities.
We do not worry about matching at infinity because we abandon the 
requirement of asymptotic flatness. It is possible that in some models of 
 bigravity such local matching does not exist  because of the 
explicit or implicit presence of ghost modes in the theory.  Such  
 models  would be  physically unacceptable. We note in this respect 
  that  the 6-dimensional model 
discussed in  \cite{Kogan:2001yr} which does not  contain negative tension branes,
 contains instead either  branes with 
equations of state violating the weak energy condition $T_{\mu\nu}^{\rm brane} \, 
\ell^{\mu} \, \ell^{\nu} \geq 0$ ( with light-like $\ell^{\mu}$) or has a conifold 
singularity in the bulk. The physical consistency of this model must be further 
investigated.  We have also quoted 
mathematical theorems linking the existence of  a hierarchical  spectrum 
 (necessary  for the derivation of an effective bigravity Lagrangian) to the necessary 
negativity of the Ricci curvature of the compactified manifold. This sign condition 
might hide the presence of ghost-like fields in the theory. These questions are  pressing
 and  deserve detailed investigation.

Assuming a positive resolution of these issues or simply taking the phenomenological 
viewpoint that nonlinear bigravity Lagrangians open an interesting new arena for non 
standard gravitational effects, we shall explore in future publications \cite{DKP1}, 
\cite{DKP2} the nonlinear physics of bigravity actions, with a particular view on its 
cosmological aspects, as it may provide a natural candidate for some new type of ``dark 
energy''.

\vskip1cm

\textbf{Acknowledgments:}
We would like to thank P. B\'{e}rard, M. Berger, A. Connes, J. Fr\"{o}hlich, M. Gromov, 
 M. Kontsevitch, A. Papazoglou, G. Ross and A. Vainshtein  for  informative  discussions.
I.K. is supported in part by PPARC 
rolling grant PPA/G/O/1998/00567 and  EC TMR grants
HPRN-CT-2000-00152 and  HRRN-CT-2000-00148.

\section*{Appendix}

In this Appendix we check the consistency of the linearized limit of the nonlinear 
action (\ref{eq6}) with a direct linearized analysis of the coupling strengths of 
massless and light graviton modes in brane models. Omitting the tensor structure (and 
considering only the relative coefficients between the various terms) the Lagrangian 
describing the coupling of the massless graviton mode $h^0$, and of the lightest one 
$h^1$, reads
\bea
\label{oldlin}
L_{Lin} = h^{0} \partial^2 h^{0} +  h^{1}\left(\partial^2 + m_1^2\right)
h^{1} + \frac{1}{M_p}\left( h^0 + \alpha_L h^1\right) T_L +
\frac{1}{M_p}\left( h^0 - \alpha_R h^1\right) T_R
\eea
where the  coefficients $\alpha_{L,R}$ describe the  relative strengths 
of the massive graviton 
coupling to the matter on left and right branes. It seems that there are four 
parameters here: $M_p,\,  m_1, \, \alpha_L,$ and $\alpha_R$. 
\begin{figure}
\begin{center}
\epsfxsize=5in
\epsfbox{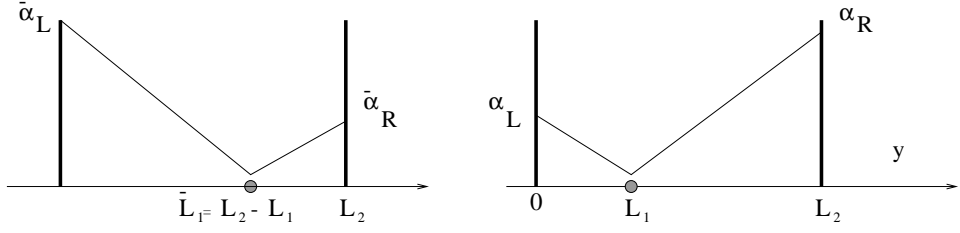}
\caption{Here $L_1$ is the position of the bounce. The left configuration is 
just the mirror image of the right one and the positions of the bounces 
are related by $\bar{L}_1 = L_2 -L_1$. Under this  transformation 
left and right branes are exchange their roles and, at the same time, 
$\alpha_R =\exp(2L_1-L_2) \rightarrow \bar{\alpha}_R = \exp(2L_2 - L_1) = 
\alpha^{-1}_R$.}
\end{center}
\end{figure}
\noindent But actually $\alpha_L = \alpha_R^{-1}$ which is extremely important 
as we shall see next. This relation  follows from the expression for $\alpha_R$ which 
was obtained in \cite{Kogan:2000wc} (see Eq.(20) and Eq.(22) there) 
\bea
\alpha_R = \exp\left(2L_1 - L_2\right) \, .
\eea
Being derived  originally for the $+-+$ model this  expression holds for other
 models with bigravity, for example  the $++$ model. In Figure 3 it is shown that 
one can interchange left and right branes by changing the position of the bounce
from $L_1$ to $\bar{L}_1 = L_2 - L_1$. One gets the new coupling strength 
\bea
\bar{\alpha}_R = \exp\left(2\bar{L}_1-L_2\right) =
\exp\left(L_2 -  2L_1\right) = \frac{1}{\alpha_R} \, .
\eea
At the same time it is easy to see that a new right brane is just an old left one, 
so that we have the result 
\bea
\alpha_L = \bar{\alpha}_R =  \frac{1}{\alpha_R} \, .
\eea
This relation is  crucial to the consistency of the nonlinear bigravity approach 
because only in this case can one  relate
 $h_0$ and $h_1$ to $h_L \sim \left ( h^0 + \alpha_L h_1\right)$ and
  $h_R \sim \left ( h^0 - \alpha_R h_1\right)$ by orthogonal rotation. If it 
were not  the case one would  get mixing between $h_L$ and $h_R$ even in the 
limit  $m_1 =0$ and we could not have two non-interacting worlds.
Introducing 
\bea
h_L = \cos\theta h^0 + \sin\theta h^1, \,\,\,,
h_R = \sin\theta h^0 - \cos\theta h^1 \,\,\,,
\tan\theta = \alpha_R
\eea
we can rewrite (\ref{oldlin}) as   
\bea
\label{linbiaction}
L_{Lin} = 
h_L \partial^2 h_L +  h_R \partial^2 h_R + \frac{m_1^2}{M_L^2 + M_R^2} 
\left(M_Lh_R - M_R h_L \right)^2 +
\frac{1}{M_L} h_L T_L + \frac{1}{M_R} h_R T_R 
\eea
where $M_L = M_p/\sqrt{1 + \alpha^2},\,\, M_R = \alpha M_p/\sqrt{1 + \alpha^2}$

Let us note that in  the limit $L_2 \rightarrow \infty$  
 both $\alpha_L$ and $M_p$ are divergent and $\alpha_R \rightarrow 0$.
In this limit $M_L$ is finite and $M_R \rightarrow \infty$.
 The massless graviton $h_R$  becomes essentially a free sterile particle
and decouples from the spectrum, while the massive 
graviton $h_R$ interacts with   matter on the left brane only.
Long range gravity completely decouples from  the  right brane. 

Here we discussed the linearized bigravity lagrangian for flat branes, 
but one can get the  same picture for $(A)dS$ branes.
The  limiting  case $L_2 \rightarrow \infty$   with corresponds to a 
single $AdS_4$ brane was considered  in \cite{Karch:2000ct} 
(see also \cite{Porrati:2001db,Porrati:2001gx,Miemiec:2000eq,Schwartz:2000ip}).

\end{document}